\PassOptionsToPackage{unicode}{hyperref}
\PassOptionsToPackage{hyphens}{url}
\PassOptionsToPackage{dvipsnames,svgnames,x11names}{xcolor}
\documentclass[
  english,
  11pt,
  a4paper,
  DIV=11,
  numbers=noendperiod]{scrartcl}
\usepackage{xcolor}
\usepackage[margin=1in]{geometry}
\usepackage{amsmath,amssymb}
\usepackage{iftex}
\ifPDFTeX
  \usepackage[T1]{fontenc}
  \usepackage[utf8]{inputenc}
  \usepackage{textcomp} 
\else 
  \usepackage{unicode-math} 
  \defaultfontfeatures{Scale=MatchLowercase}
  \defaultfontfeatures[\rmfamily]{Ligatures=TeX,Scale=1}
\fi
\usepackage{lmodern}
\ifPDFTeX\else
\fi
\IfFileExists{upquote.sty}{\usepackage{upquote}}{}
\IfFileExists{microtype.sty}{
  \usepackage[]{microtype}
  \UseMicrotypeSet[protrusion]{basicmath} 
}{}
\makeatletter
\@ifundefined{KOMAClassName}{
  \IfFileExists{parskip.sty}{%
    \usepackage{parskip}
  }{
    \setlength{\parindent}{0pt}
    \setlength{\parskip}{6pt plus 2pt minus 1pt}}
}{
  \KOMAoptions{parskip=half}}
\makeatother
\makeatletter
\ifx\paragraph\undefined\else
  \let\oldparagraph\paragraph
  \renewcommand{\paragraph}{
    \@ifstar
      \xxxParagraphStar
      \xxxParagraphNoStar
  }
  \newcommand{\xxxParagraphStar}[1]{\oldparagraph*{#1}\mbox{}}
  \newcommand{\xxxParagraphNoStar}[1]{\oldparagraph{#1}\mbox{}}
\fi
\ifx\subparagraph\undefined\else
  \let\oldsubparagraph\subparagraph
  \renewcommand{\subparagraph}{
    \@ifstar
      \xxxSubParagraphStar
      \xxxSubParagraphNoStar
  }
  \newcommand{\xxxSubParagraphStar}[1]{\oldsubparagraph*{#1}\mbox{}}
  \newcommand{\xxxSubParagraphNoStar}[1]{\oldsubparagraph{#1}\mbox{}}
\fi
\makeatother

\usepackage{longtable,booktabs,array}
\usepackage{calc} 
\usepackage{etoolbox}
\makeatletter
\patchcmd\longtable{\par}{\if@noskipsec\mbox{}\fi\par}{}{}
\makeatother
\IfFileExists{footnotehyper.sty}{\usepackage{footnotehyper}}{\usepackage{footnote}}
\makesavenoteenv{longtable}
\usepackage{graphicx}
\makeatletter
\newsavebox\pandoc@box
\newcommand*\pandocbounded[1]{
  \sbox\pandoc@box{#1}%
  \Gscale@div\@tempa{\textheight}{\dimexpr\ht\pandoc@box+\dp\pandoc@box\relax}%
  \Gscale@div\@tempb{\linewidth}{\wd\pandoc@box}%
  \ifdim\@tempb\p@<\@tempa\p@\let\@tempa\@tempb\fi
  \ifdim\@tempa\p@<\p@\scalebox{\@tempa}{\usebox\pandoc@box}%
  \else\usebox{\pandoc@box}%
  \fi%
}
\def\fps@figure{htbp}
\makeatother

\NewDocumentCommand\citeproctext{}{}

\makeatletter
 \let\@cite@ofmt\@firstofone
 \def\@biblabel#1{}
 \def\@cite#1#2{{#1\if@tempswa , #2\fi}}
\makeatother
\newlength{\cslhangindent}
\newlength{\csllabelwidth}
\newenvironment{CSLReferences}[2] 
 {\begin{list}{}{%
  \setlength{\itemindent}{0pt}
  \setlength{\leftmargin}{0pt}
  \setlength{\parsep}{0pt}
  \ifodd #1
   \setlength{\leftmargin}{\cslhangindent}
   \setlength{\itemindent}{-1\cslhangindent}
  \fi
  \setlength{\itemsep}{#2\baselineskip}}}
 {\end{list}}
\usepackage{calc}

\ifLuaTeX
\usepackage[bidi=basic,shorthands=off]{babel}
\else
\usepackage[bidi=default,shorthands=off]{babel}
\fi
\ifLuaTeX
  \usepackage{selnolig} 
\fi

\usepackage{booktabs}
\usepackage{longtable}
\usepackage{array}
\usepackage{multirow}
\usepackage{wrapfig}
\usepackage{float}
\usepackage{colortbl}
\usepackage{pdflscape}
\usepackage{tabularx}
\usepackage{xltabular}
\usepackage{threeparttable}
\usepackage{threeparttablex}
\usepackage[normalem]{ulem}
\usepackage{makecell}
\usepackage{xcolor}
\usepackage[T1]{fontenc}
\AtBeginDocument{
  \counterwithout{proposition}{section}
  \counterwithout{corollary}{section}
}
\KOMAoption{captions}{tableheading}
\usepackage{needspace}
\makeatletter
\@ifpackageloaded{caption}{}{\usepackage{caption}}
\AtBeginDocument{%
\ifdefined\contentsname
  \renewcommand*\contentsname{Table of contents}
\else
  \newcommand\contentsname{Table of contents}
\fi
\ifdefined\listfigurename
  \renewcommand*\listfigurename{List of Figures}
\else
  \newcommand\listfigurename{List of Figures}
\fi
\ifdefined\listtablename
  \renewcommand*\listtablename{List of Tables}
\else
  \newcommand\listtablename{List of Tables}
\fi
\ifdefined\figurename
  \renewcommand*\figurename{Figure}
\else
  \newcommand\figurename{Figure}
\fi
\ifdefined\tablename
  \renewcommand*\tablename{Table}
\else
  \newcommand\tablename{Table}
\fi
}
\@ifpackageloaded{float}{}{\usepackage{float}}
\floatstyle{ruled}
\@ifundefined{c@chapter}{\newfloat{codelisting}{h}{lop}}{\newfloat{codelisting}{h}{lop}[chapter]}
\floatname{codelisting}{Listing}

\usepackage{amsthm}
\theoremstyle{plain}
\newtheorem{corollary}{Corollary}[section]
\theoremstyle{plain}
\newtheorem{proposition}{Proposition}[section]
\theoremstyle{remark}
\AtBeginDocument{}

\makeatother
\makeatletter
\@ifpackageloaded{caption}{}{\usepackage{caption}}
\@ifpackageloaded{subcaption}{}{\usepackage{subcaption}}
\makeatother
\usepackage{bookmark}
\IfFileExists{xurl.sty}{\usepackage{xurl}}{} 
\makeatletter
\@ifundefined{xmpquote}{\newcommand{\xmpquote}[1]{#1}}{}
\makeatother
\hypersetup{
  pdftitle={When Do Type-Specific Wages Buffer Distributional Incidence in TANK?},
  pdfauthor={Kenji Miyazaki},
  pdflang={en},
  pdfkeywords={\xmpquote{TANK, type-specific wages, labor reallocation,
cross-type consumption gap, monetary transmission, distributional
incidence}},
  colorlinks=true,
  linkcolor={blue},
  filecolor={Maroon},
  citecolor={Blue},
  urlcolor={Blue},
  pdfcreator={LaTeX via pandoc}}

\title{When Do Type-Specific Wages Buffer Distributional Incidence in
TANK?}
\author{Kenji Miyazaki\footnote{Faculty of Economics, Hosei University;
  \texttt{miya\_ken@hosei.ac.jp}. The author thanks Daisuke Ida, Hiroshi
  Gunji, and Kazuki Hiraga for useful comments and suggestions and
  acknowledges financial support from the Ministry of Education,
  Science, Sports, and Culture through Grants-in-Aid for Scientific
  Research (C), grant numbers 23K01384 and 26K04923.}}
\date{2026-09-22}
\begin{document}
\maketitle
\begin{abstract}
When do relative wages buffer the unequal incidence of aggregate shocks?
I derive a consumption-gap decomposition and a present-value condition
for partial offset in a TANK model. An extension separates wage-setting
demand elasticity from substitution between labor segments and allows
each segment to contain both financial types. With a zero inherited wage
gap and a same-sign discounted wedge, substitution above one gives
offsetting earnings reallocation; substitution below one gives
amplification. The channel disappears when financial types have
identical segment exposure. Numerical experiments assess these
mechanisms, shock persistence, policy feedback, and aggregate-IRF
matching. In the nested perfect-alignment monetary benchmark, the peak
consumption gap is about two-fifths smaller under type-specific wages
than under the common-wage closure. These are conditional model
comparisons, not empirical effect estimates or welfare rankings.
\end{abstract}



\section{Introduction}\label{introduction}

Aggregate shocks redistribute consumption through profits, transfers,
and labor income. A common-wage TANK closure rules out one margin of
this incidence: relative wages cannot reallocate employment and earnings
between hand-to-mouth households and savers. This paper asks whether a
common-wage closure can approximate both aggregate and distributional
responses when the two groups instead supply imperfectly substitutable
labor and their wages adjust separately. I show when relative earnings
offset unequal profit-transfer exposure and why the adequacy of the
approximation can depend on the outcome of interest.

In the tractable TANK economy, hand-to-mouth households consume current
disposable income; savers trade bonds and receive dividends. The
benchmark deliberately aligns financial types with labor-market
segments: each group supplies its own imperfectly substitutable labor
type. This benchmark restriction ties changes in relative labor income
to the same groups that differ in asset-market access; the labor-market
extension relaxes that alignment. A higher relative wage for one type
reduces its relative employment. When the elasticity of substitution
exceeds one, the employment response dominates the wage component of
relative earnings. Relative labor income can then move in the opposite
direction from unequal profit-transfer incidence. The term
\emph{distributional buffer} denotes this partial offset in consumption
incidence and does not imply a welfare improvement.

Within the log-linearized model, the central analytical result is an
exact decomposition of the signed consumption gap into direct
profit-transfer incidence and a relative-earnings term. With a zero
inherited wage gap and nonzero current incidence, strict partial offset
is equivalent to an explicit bound on the discounted expected wedge
relative to its current value. I derive the wedge from the aggregate
real-wage equation and evaluate that condition in the full equilibrium.
Under the same zero-inherited-gap condition, an AR(1)-type conditional
wedge path guarantees partial offset, but this path restriction is not
imposed on the benchmark equilibrium. The channel also remains operative
in the flexible type-specific-wage limit, so nominal wage rigidity is
not required for the buffer itself.

Own-lag wage adjustment adds an inherited wage-gap state. The stable
solution separates its contribution from expected equilibrium-wedge
forcing, distinguishing the persistence due to inherited wages from that
due to the expected path of direct incidence. This decomposition
explains the dynamics of the channel without making nominal rigidity a
prerequisite for earnings reallocation.

The extension distinguishes substitution between labor segments from the
demand elasticity facing individual wage setters. With zero inherited
relative wages and a discounted wedge of the same sign as current
incidence, substitution above one generates offsetting relative
earnings; substitution below one generates amplification. A
mixed-segment union environment also allows each labor segment to
include both financial types. Identical segment composition removes the
relative-earnings channel between financial types. These results
identify the assumptions behind the buffer rather than interpreting
nominal wage rigidity itself as sufficient for attenuation.

The monetary experiment illustrates why this mechanism matters when
comparing the two closures. At the nested perfect-alignment benchmark,
the peak consumption gap is about two-fifths smaller with type-specific
wages, while the peak output responses are close. This contrast is
outcome specific: the common-wage approximation closely tracks the
benchmark peak output response but fails to capture much of the
distributional incidence. The comparison includes general-equilibrium
adjustment and does not identify the effect of a single parameter.

The benchmark equates the two labor elasticities to recover the original
single-CES model; it does not independently calibrate substitution
between labor segments. Partial alignment and lower segment substitution
substantially reduce the consumption-gap difference. Additional
comparisons fit the common-wage model to aggregate output and inflation
responses before evaluating its consumption gap. The size of the buffer
is therefore a conditional calibration result, whereas the extension
characterizes when its sign changes or the channel disappears.

The quantitative analysis distinguishes accounting offsets within a
given equilibrium from consumption-gap reductions across closures. It
evaluates impact incidence, the subsequent response path, and
sensitivity to labor-market structure, policy, and shock dynamics.
Similar aggregate responses in the monetary benchmark do not imply
uniformly close responses across these experiments.

Tractable heterogeneous-agent models isolate how limited asset-market
participation, profits, transfers, and marginal propensities to consume
shape aggregate demand and consumption dispersion (Bilbiie 2008, 2020,
2025). Debortoli and Galí (2025) show that a suitably specified TANK
model can reproduce many aggregate implications of richer HANK models.
Kaplan et al. (2018) emphasize that much of monetary transmission to
consumption operates indirectly through general-equilibrium labor-income
effects. The approximation question here is instead whether replacing
type-specific wage setting with a common-wage closure preserves
cross-type incidence as well as aggregate transmission. The benchmark
finding that output gaps are similar while consumption gaps differ adds
an outcome-specific qualification to comparisons based on aggregate
dynamics alone.

In the literature on distributional mechanisms, Auclert (2019)
identifies earnings heterogeneity as one channel through which unequal
income gains interact with heterogeneous marginal propensities to
consume. Komatsu (2023) examines consumption inequality in a wage-rigid
TANK model with search-and-matching frictions. In that model, the
earnings-heterogeneity channel operates through unemployment and
matching. Here, the relevant margin is relative labor demand rather than
unemployment. Even without search frictions, type-specific wages
reallocate employment and earnings between the same two groups that
differ in asset-market access.

This mechanism complements research on aggregate dynamics in sticky-wage
TANK models (Furlanetto 2011; Colciago 2011; Ascari et al. 2017; Ida and
Okano 2024) and work on the labor-supply and earnings block in
sticky-wage HANK models (Auclert et al. 2023; Gerke et al. 2024). Gerke
et al. (2024) contrast the standard homogeneous-hours allocation with
rules that assign different hours across heterogeneous households. In
contrast, the present model assigns separate wages to two labor segments
and aligns each segment with a financial type.

More directly related to the wage-rigidity question, Diz et al. (2023)
study a common-wage TANK model in which hand-to-mouth disposable income
consists only of labor income. They show that greater aggregate wage
flexibility relative to sticky prices can amplify output fluctuations
following demand shocks. Bojovic and Ellis (2026) show that this
conclusion can reverse when hand-to-mouth households have lower
productivity, receive an allocation of firm profits, and face
debt-service exposure. Both analyses vary aggregate wage rigidity
without introducing financial-type-specific wages. The present paper
instead asks which aspects of cross-type incidence that restriction
misses when relative wages can reallocate employment and earnings across
financial types.

The cross-closure result is also related to work that separates
aggregate allocations from factor-income incidence. Broer et al. (2020)
show in a tractable HANK model that profit and labor-supply interactions
make wage rigidity consequential for monetary transmission. Bilbiie and
Trabandt (2025) establish an aggregate equivalence between polar
sticky-price and sticky-wage representative-agent economies even though
wages and profits differ. They show that the equivalence generally
breaks with household heterogeneity.

The present comparison differs from that polar equivalence: both
closures retain sticky prices, and the common-wage case retains an
aggregate sticky-wage equation. The narrower point is that a common-wage
closure can track selected aggregate dynamics while suppressing the
relative-wage earnings channel. This is related only in spirit to the
approximate aggregation result of Krusell and Smith (1998), which
concerns mean wealth in a flexible-price incomplete-markets economy
rather than the wage-setting closures considered here. The specific
contribution here is a closed-form relative-wage accounting under
imperfect substitutability between labor types, together with a
separately quantifiable own-lag state.

The closest companion paper is Miyazaki (2026), which studies profit
redistribution and aggregate monetary transmission in a TANK model using
the same core ingredients: type-specific wages, imperfectly
substitutable labor, and unequal profit exposure. Its closed-form
benchmark specifies each type's wage adjustment relative to the same
lagged aggregate wage---interpreted as a reduced-form external wage
norm---and then compares that aggregate-lag economy with conventional
own-lag adjustment while holding primitive adjustment-cost coefficients
fixed. The aggregate-lag benchmark preserves the relative-wage
reallocation of employment and earnings, but it does not create the
inherited wage-gap state that arises when each type's current wage
becomes its own reference next period. The present paper instead takes
the own-lag model as its maintained environment and shifts the focus
from adjustment-cost timing and aggregate amplification to cross-type
consumption incidence: it derives the consumption-gap decomposition,
compares type-specific wages with a common-wage closure, separates the
inherited state from expected distributional-wedge forcing, and
distinguishes between-segment substitution from financial-type
alignment. Accordingly, Table~\ref{tbl-wage-specifications} lists only
the two wage-setting environments analyzed in the present closure
comparison rather than treating the aggregate-lag model as a third
specification.

The mixed-segment extension identifies a first-order common-wage
endpoint. Table~\ref{tbl-companion-scope} distinguishes the principal
analytical objects without treating the shared mechanism as new in each
paper.

\Needspace{0.46\textheight}

\begin{longtable}[]{@{}
  >{\raggedright\arraybackslash}p{(\linewidth - 4\tabcolsep) * \real{0.1800}}
  >{\raggedright\arraybackslash}p{(\linewidth - 4\tabcolsep) * \real{0.4000}}
  >{\raggedright\arraybackslash}p{(\linewidth - 4\tabcolsep) * \real{0.4200}}@{}}
\caption{Relationship to the companion
paper}\label{tbl-companion-scope}\tabularnewline
\toprule\noalign{}
\begin{minipage}[b]{\linewidth}\raggedright
Dimension
\end{minipage} & \begin{minipage}[b]{\linewidth}\raggedright
Companion paper
\end{minipage} & \begin{minipage}[b]{\linewidth}\raggedright
Present paper
\end{minipage} \\
\midrule\noalign{}
\endfirsthead
\toprule\noalign{}
\begin{minipage}[b]{\linewidth}\raggedright
Dimension
\end{minipage} & \begin{minipage}[b]{\linewidth}\raggedright
Companion paper
\end{minipage} & \begin{minipage}[b]{\linewidth}\raggedright
Present paper
\end{minipage} \\
\midrule\noalign{}
\endhead
\bottomrule\noalign{}
\endlastfoot
Main question & Aggregate amplification and wage-adjustment timing &
Distributional incidence and closure approximation \\
Primary comparison & Aggregate-lag and own-lag wage adjustment &
Type-specific and common-wage incidence \\
Analytical emphasis & Multipliers, subsidy financing, welfare & Gap
accounting, expected-wedge criterion, inherited state \\
Additional structure here & Shared wage and profit mechanisms & Separate
labor elasticities and mixed financial types \\
\end{longtable}

\emph{Notes:} The papers share a model lineage and some benchmark
mechanisms. The comparison identifies their principal analytical
objects; it does not imply that the companion paper excludes all
common-wage or distributional analysis.

Section 2 derives the consumption-gap identity, the wage-gap law, and
the reduced-form common-wage closure. Section 3 derives the general
offset criterion, separates inherited dynamics from expected-wedge
forcing, and extends the labor-market structure. Section 4 reports
within-path accounting, aggregate-matched closure comparisons, and
sensitivity to substitution, alignment, and shock dynamics. Section 5
concludes.

\section{Model and Canonical
Representation}\label{model-and-canonical-representation}

The model isolates a simple incidence channel in a TANK economy with
sticky prices and type-specific sticky wages. Hand-to-mouth households
consume current disposable income, while savers trade bonds and receive
firm dividends. Because the two groups supply imperfectly substitutable
labor, relative wages reallocate employment and earnings between the
same household groups that differ in profit exposure. When this
earnings-reallocation term offsets direct incidence, it acts as the
distributional buffer; own-lag wage adjustment adds a separate inherited
wage-gap state.

This section presents the baseline environment, states the
log-linearized equilibrium system, and derives the canonical reduction.
The online appendix presents the nonlinear household, labor-packer,
firm, and government problems, together with the first-order
log-linearization around the symmetric zero-inflation steady state.

Throughout, lower-case letters denote log deviations from the symmetric
zero-inflation steady state, as in \(c_t = \log(C_t/C)\). The aggregate
profit/dividend share \(d_t=\mathcal D_t/Y\) and per-Type-H-household
transfer \(z_t=Z_t/Y\) are exceptions: both are normalized by
steady-state output \(Y\), not logged. Aggregate transfers equal
\(\lambda z_tY\). The baseline dynamic equations assume
\(\eta_p,\eta_w>0\); zero adjustment costs are treated through
separately derived flexible-price or flexible-wage limits. For any
variable \(q_t\), \(\Delta q_{t+1}\equiv q_{t+1}-q_t\) denotes its first
difference. Let \(\lambda\in(0,1)\) denote the population share of Type
H households. For \(q\in\{c,w,n\}\), define the population-weighted
aggregate and signed cross-type gap by \[
q_t=\lambda q_t^H+(1-\lambda)q_t^S,
\qquad
\sigma_t^q=q_t^S-q_t^H.
\] Here \(c\), \(w\), \(n\), and \(a\) denote log consumption, the log
real wage, log labor input, and log technology, respectively.
Accordingly, \(\sigma_t^c\), \(\sigma_t^w\), and \(\sigma_t^n\) are
signed Type-S-minus-Type-H gaps rather than variance measures;
\(\sigma_t^n\) is a labor-input gap, not an unemployment rate.

\subsection{Environment}\label{environment}

\textbf{Households.}

Households form a continuum of measure one. Type H households have
population share \(\lambda \in (0, 1)\), and Type S households have
share \(1-\lambda\). Both types have the same period utility, with
discount factor \(\beta \in (0,1)\), inverse elasticity of intertemporal
substitution \(\gamma > 0\), and inverse Frisch elasticity
\(\varphi > 0\). Type S households trade one-period nominal bonds and
receive firm dividends. Type H households do not participate in asset
markets.

The benchmark deliberately aligns financial type with labor-market
segment; Section~\ref{sec-segmentation-main} subsequently relaxes that
restriction. This one-to-one mapping is a maintained modeling
restriction: it makes relative labor demand reallocate earnings across
the same two groups that differ in asset-market access. It should not be
read as a claim that liquidity status is itself an occupation or
bargaining unit. An empirical application would need to discipline the
correlation between household liquidity status and the wage-setting
segments that face different labor demand.

Let \(W_t^j \equiv W_t^{N,j}/P_t\) denote the type-specific real wage
for \(j \in \{H,S\}\). Aggregate price inflation and type-specific wage
inflation are defined by \[
\Pi_t^p \equiv \frac{P_t}{P_{t-1}}, \qquad
\Pi_t^j \equiv \frac{W_t^{N,j}}{W_{t-1}^{N,j}}, \quad j \in \{H,S\}.
\]

\textbf{Labor market and wage setting.}

The labor market has differentiated labor services and nominal wage
stickiness. Households supply labor varieties to a representative labor
packer, which aggregates them into a homogeneous labor input. The
baseline elasticity of substitution across labor varieties is
\(\psi_w>1\), and wage setting is subject to Rotemberg adjustment costs
with \(\eta_w>0\). The extension in Section~\ref{sec-segmentation-main}
separates within-segment demand elasticity from between-segment
substitution. The same wage-adjustment coefficient applies to both
types. Each household type adjusts its nominal wage relative to its own
previous nominal wage, rather than relative to an aggregate wage index.

Under within-type symmetry, the competitive labor packer combines the
two labor types according to \[
N_t=\left[\lambda (N_t^H)^{\frac{\psi_w-1}{\psi_w}}+(1-\lambda)(N_t^S)^{\frac{\psi_w-1}{\psi_w}}\right]^{\frac{\psi_w}{\psi_w-1}},
\] with the nominal wage index \[
W_t^N=\left[\lambda (W_t^{N,H})^{1-\psi_w}+(1-\lambda)(W_t^{N,S})^{1-\psi_w}\right]^{\frac{1}{1-\psi_w}}.
\] Cost minimization gives the type-specific labor demand schedule \[
N_t^i=N_t\left(\frac{W_t^{N,i}}{W_t^N}\right)^{-\psi_w},
\qquad i\in\{H,S\}.
\] Thus, using the signed-gap notation above, the first-order relative
labor-demand and labor-income gaps satisfy \[
\sigma_t^n=-\psi_w\sigma_t^w,
\qquad
\sigma_t^{w+n}\equiv\sigma_t^w+\sigma_t^n=(1-\psi_w)\sigma_t^w.
\] The population share \(\lambda\) weights the two labor varieties; it
is not a fixed input requirement. Relative wages therefore reallocate
employment endogenously between household types. This
earnings-reallocation channel arises from type-specific wages and
imperfect substitutability between the two labor types. Nominal wage
rigidity is not required for the channel to operate: relative wages,
employment, and earnings continue to adjust in the flexible-wage limit
derived below. Own-lag wage adjustment instead alters the dynamics of
the channel by introducing a forward-looking wage-setting term and an
inherited wage-gap state, \(\sigma_{t-1}^w\).

The own-lag specification uses each segment's previous nominal wage as
the reference point for renegotiation. It is a reduced-form
representation of segment-specific contracts, not evidence that
hand-to-mouth status determines the bargaining unit.\footnote{ Wage
  changes are often governed by contract-, firm-, industry-,
  occupation-, or skill-specific arrangements rather than by a
  continuously adjusted aggregate wage. Cross-country evidence links
  wage-setting timing to monetary transmission (Olivei and Tenreyro
  2010). Agreement data document industry- and firm-level wage rigidity
  in France (Avouyi-Dovi et al. 2013), fixed-wage contracts in Sweden
  (Björklund et al. 2019), and province-sector-skill-specific wage
  floors in Spain (Adamopoulou et al. 2024). These findings motivate
  segment-specific own-lag contracts, but the alignment between contract
  segment and financial type remains a model assumption.} Because
today's wage becomes the household's own-lag reference in the next
period, the adjustment-cost problem jointly generates the
forward-looking term and the inherited state. The maintained model does
not treat these terms as independently switchable primitives.

Table~\ref{tbl-wage-specifications} summarizes the two specifications
used for the main closure comparison. The baseline is derived from the
household wage-setting problem with own-lag adjustment costs. The
common-wage closure instead removes the Type H wage-setting block and
the entire relative-wage and employment-reallocation margin while
retaining aggregate sticky-wage dynamics. It is an explicit reduced-form
benchmark rather than a second adjustment-cost model or a restriction
nested through \(\eta_w\). Appendix A.3 gives its algebraic reduction.
The mixed-segment extension in Section~\ref{sec-segmentation-main}
supplies a separate union environment whose identical-composition
endpoint reproduces this closure at first order.

\begin{table}[H]

\caption{\label{tbl-wage-specifications}Wage setting in the main closure
comparison}

\centering{

\centering\begingroup\fontsize{10}{12}\selectfont

\begin{tabular}[t]{>{\raggedright\arraybackslash}p{45mm}>{\raggedright\arraybackslash}p{42mm}>{\centering\arraybackslash}p{27mm}>{\centering\arraybackslash}p{27mm}}
\toprule
Specification & Role in analysis & Cross-type employment reallocation? & Inherited wage-gap state?\\
\midrule
Type-specific own-lag Rotemberg wage setting & Baseline adjustment-cost model & Yes & Yes\\
Common-wage closure & Reduced-form common-wage benchmark & No & No\\
\bottomrule
\end{tabular}
\endgroup{}

}

\end{table}%

The first row describes the maintained model used for the analytical and
quantitative results; its forward-looking term and inherited state are
jointly derived from the own-lag adjustment-cost problem. The second row
is the reduced-form benchmark used to measure what is lost when relative
wages cannot reallocate employment and earnings. In the original
household wage-setting model it is not obtained by varying the
adjustment-cost parameter; in the mixed-segment union extension,
identical financial-type composition reproduces its first-order
aggregate and consumption equations.

\textbf{Terminology and analytical objects.}

Throughout, \(\sigma_t^w=w_t^S-w_t^H\) denotes the signed cross-type
real-wage gap, referred to below as the wage gap; it is distinct from
nonnegative wage-dispersion indices associated with staggered nominal
adjustment. As shown below, the net distributional wedge is
\(\omega_t=d_t-z_t=a_t-w_t-z_t\), and the cross-type consumption gap
satisfies \(\sigma_t^c=\omega_t/(1-\lambda)+(1-\psi_w)\sigma_t^w\). I
call \((1-\psi_w)\sigma_t^w\) the \emph{earnings-reallocation term} and
refer to it as a \emph{buffer} when it offsets the direct-incidence term
\(\omega_t/(1-\lambda)\). Under own-lag adjustment, \(\sigma_{t-1}^w\)
is the inherited wage-gap state; the current wage gap combines the
contribution from that state with expected equilibrium-wedge forcing.
The reduced-form common-wage closure imposes \(w_t^H=w_t^S=w_t\) and
removes the wage-gap law and associated employment-reallocation margin
while retaining aggregate sticky-wage dynamics.

\textbf{Firms and price setting.}

The production sector comprises a final-goods aggregator and a continuum
of intermediate firms. Intermediate firms produce differentiated goods
with linear technology \(Y_t(\ell)=\exp(a_t)N_t(\ell)\), where \(a_t\)
is a neutral technology shock. They set prices subject to Rotemberg
adjustment costs with parameter \(\eta_p>0\). Aggregate real dividends
\(\mathcal D_t\) accrue exclusively to Type S households, so each saver
receives \(D_t^S=\mathcal D_t/(1-\lambda)\). Constant wage and sales
subsidy rates remove the steady-state markups. Each subsidy is financed
by a contemporaneous lump-sum levy on its recipient group, so its level
does not generate a separate net-income transfer in equilibrium. The
rates are constant, while expenditure and the matching levies vary with
earnings and sales.

\textbf{Aggregation, transfers, and monetary policy.}

\nopagebreak[4]

At first order, aggregate consumption, labor, and real wages follow the
population-weighted convention defined at the start of this section. The
online appendix presents the household problems and the full
labor-packer derivation.

Let \(Z_t\) denote the transfer received by each Type H household: \[
Z_t = z_t Y,
\] where \(Y\) is steady-state output and \(z_t\) is a normalized
per-recipient transfer. Total transfers are \(\lambda Z_t\), financed by
a levy \(\lambda Z_t/(1-\lambda)\) on each saver. A fixed \(z_t\)
therefore implies a larger aggregate fiscal intervention when
\(\lambda\) rises. The quantitative analysis distinguishes
fixed-per-recipient and fixed-total-transfer experiments.

In real units, the fiscal liabilities that also finance the subsidies
are
\begin{equation}\protect\phantomsection\label{eq-fiscal-liabilities}{
\begin{aligned}
T_t^H&=-Z_t+\tau_wW_t^HN_t^H, &
T_t^S&=\frac{\lambda}{1-\lambda}Z_t+\tau_wW_t^SN_t^S,\\
T_t^M&=\tau_pY_t, &
\tau_w&=\frac{1}{\psi_w-1},\qquad \tau_p=\frac{1}{\psi_p-1}.
\end{aligned}
}\end{equation} Consequently, total fiscal revenue equals wage- and
sales-subsidy expenditure in every period:
\begin{equation}\protect\phantomsection\label{eq-government-budget}{
\lambda T_t^H+(1-\lambda)T_t^S+T_t^M
=\tau_w[\lambda W_t^HN_t^H+(1-\lambda)W_t^SN_t^S]+\tau_pY_t.
}\end{equation} Individual households and firms take these lump-sum
liabilities as given when choosing wages and prices. Substituting the
levies into their budgets cancels the subsidy receipts in equilibrium,
leaving the intended transfer between financial types. Net dividends are
output less the wage bill and price-adjustment costs; hence
\(d_t=y_t-w_t-n_t=a_t-w_t\) at first order. A positive \(Z_t\) makes the
transfer component of \(T_t^H\) negative, but need not make the entire
fiscal liability negative.

Monetary policy follows \[
R_t^N = \frac{\exp(-m_t)}{\beta}(\Pi_t^p)^{\phi_\pi},
\] where \(\phi_\pi>1\) is the response to price inflation and \(m_t\)
is the monetary policy shock. In the first-order system, the ex ante
real rate is \[
r_t = r_t^N - \mathbb{E}_t \pi_{t+1}^p,
\] where \(r_t^N\) and \(\pi_t^p\) are the log deviations of \(R_t^N\)
and \(\Pi_t^p\), respectively.

Rotemberg price setting, type-specific Rotemberg wage setting with
own-lag wage inflation, and the redistributive transfer disturbance
\(z_t\) define the dual-rigidity TANK environment. The transfer
disturbance follows an exogenous AR(1) process. The term \emph{passive
transfers} denotes this exogenous process, with no feedback that targets
the distributional wedge.

I also consider a reduced-form common-wage closure related to
homogeneous-labor-supply arrangements in sticky-wage HANK models (Gerke
et al. 2024). The closure imposes \(w_t^H=w_t^S=w_t\), removes the Type
H wage-setting block, and retains the aggregate sticky-wage equation.
Within the original household wage-setting model, this is a reduced-form
change of institution rather than a restriction obtained by varying
\(\eta_w\). Section~\ref{sec-segmentation-main} supplies a separate
union extension with a common-wage first-order endpoint. Under CES labor
demand, equal wages imply \(n_t^H=n_t^S=n_t\), so common labor supply
follows from the closure rather than from an additional restriction.

\begin{table}

\caption{\label{tbl-notation}Primitive parameters}

\centering{

\begin{tabular}[t]{ll}
\toprule
Symbol & Meaning\\
\midrule
\addlinespace[0.3em]
\multicolumn{2}{l}{\textbf{Preferences / Heterogeneity}}\\
\hspace{1em}$\beta$ & Discount factor\\
\hspace{1em}$\gamma$ & Inverse elasticity of intertemporal substitution\\
\hspace{1em}$\varphi$ & Inverse Frisch elasticity\\
\hspace{1em}$\lambda$ & Share of hand-to-mouth households\\
\addlinespace[0.3em]
\multicolumn{2}{l}{\textbf{Nominal Rigidities / Market Structure}}\\
\hspace{1em}$\psi_p$ & Elasticity of substitution across goods\\
\hspace{1em}$\eta_p$ & Rotemberg price adjustment cost\\
\hspace{1em}$\psi_w$ & Elasticity of substitution across labor types\\
\hspace{1em}$\eta_w$ & Rotemberg wage adjustment cost, common across household types\\
\addlinespace[0.3em]
\multicolumn{2}{l}{\textbf{Policy / Shock Processes}}\\
\hspace{1em}$\phi_\pi$ & Taylor-rule response to price inflation\\
\hspace{1em}$\rho_a$ & Persistence of the technology shock\\
\hspace{1em}$\rho_m$ & Persistence of the monetary shock\\
\hspace{1em}$\rho_z$ & Persistence of the transfer shock\\
\bottomrule
\end{tabular}

}

\end{table}%

\subsection{The Log-Linearized
Economy}\label{the-log-linearized-economy}

The log-linearized economy comprises a nonredundant system of 20
equations in 20 endogenous variables. This subsection presents that
system and highlights the blocks that yield the five-equation dynamic
representation below. Table~\ref{tbl-notation} lists the primitive
parameters used in the derivation.

I write \(\pi^p_t\) for price inflation, \(\pi^w_t\) for aggregate
nominal wage inflation, \(\mu^p_t\) for the price markup, and
\(\mu^w_t\) for the aggregate wage markup. Household-specific variables
with superscript \(j \in \{H,S\}\) follow the same notation.

\textbf{Aggregation and cross-type gaps.}

The canonical reduction uses the aggregate/gap definitions stated at the
start of this section. Once the aggregate level and the signed
cross-type gap are known, the allocations of Type H and Type S
households follow directly.

\textbf{The full system of equations.}

The full system consists of 20 equilibrium equations and three exogenous
shock processes. Four equations drive the mechanism: the Type H budget
constraint, the own-lag wage-inflation identity, Type H labor demand,
and the profit-share definition. Together, they link transfers, the wage
gap, the cross-type labor-income gap, and the net distributional wedge.

The Taylor rule closes the aggregate block. Table~\ref{tbl-full-system}
organizes the equations into the household, wage, price, and
market-clearing blocks used in the derivation. The online appendix
presents the nonlinear conditions, intermediate log-linearization steps,
and a production-curvature extension with \(Y_t=A_tN_t^{1-\alpha}\). For
this extension, it reports both the extended first-order system and the
corresponding canonical-form table.

\begin{table}

\caption{\label{tbl-full-system}Log-linear equilibrium system}

\centering{

\begin{tabular}[t]{ll}
\toprule
Name & Equation\\
\midrule
\addlinespace[0.3em]
\multicolumn{2}{l}{\textbf{Household Block}}\\
\hspace{1em}Type S Euler equation & $r_t = \gamma \mathbb{E}_t \Delta c^S_{t+1}$\\
\hspace{1em}Type H budget constraint & $c^H_t = w^H_t + n^H_t + z_t$\\
\hspace{1em}Aggregate consumption aggregation & $c_t = \lambda c^H_t + (1-\lambda) c^S_t$\\
\hspace{1em}Aggregate labor aggregation & $n_t = \lambda n^H_t + (1-\lambda) n^S_t$\\
\addlinespace[0.3em]
\multicolumn{2}{l}{\textbf{Wage Block}}\\
\hspace{1em}Aggregate wage Phillips curve & $\pi^w_t = \beta \mathbb{E}_t \pi^w_{t+1} - \frac{\psi_w}{\eta_w} \mu^w_t$\\
\hspace{1em}Aggregate wage-markup definition & $\mu^w_t = w_t - \gamma c_t - \varphi n_t$\\
\hspace{1em}Aggregate wage-inflation identity & $\pi^w_t = w_t - w_{t-1} + \pi^p_t$\\
\hspace{1em}Type H wage Phillips curve & $\pi^H_t = \beta \mathbb{E}_t \pi^H_{t+1} - \frac{\psi_w}{\eta_w} \mu^H_t$\\
\hspace{1em}Type H wage-markup definition & $\mu^H_t = w^H_t - \gamma c^H_t - \varphi n^H_t$\\
\hspace{1em}Type H wage-inflation identity & $\pi^H_t = w^H_t - w^H_{t-1} + \pi^p_t$\\
\hspace{1em}Type H labor demand & $n^H_t = n_t - \psi_w (w^H_t - w_t)$\\
\hspace{1em}Aggregate wage aggregation identity & $w_t = \lambda w^H_t + (1-\lambda) w^S_t$\\
\hspace{1em}Aggregate wage-markup aggregation identity & $\mu^w_t = \lambda \mu^H_t + (1-\lambda) \mu^S_t$\\
\hspace{1em}Aggregate wage-inflation aggregation identity & $\pi^w_t = \lambda \pi^H_t + (1-\lambda) \pi^S_t$\\
\addlinespace[0.3em]
\multicolumn{2}{l}{\textbf{Price Block}}\\
\hspace{1em}Price Phillips curve & $\pi^p_t = \beta \mathbb{E}_t \pi^p_{t+1} - \frac{\psi_p}{\eta_p} \mu^p_t$\\
\hspace{1em}Price-markup definition & $\mu^p_t = a_t - w_t$\\
\hspace{1em}Profit-share definition & $d_t = a_t - w_t$\\
\addlinespace[0.3em]
\multicolumn{2}{l}{\textbf{Market-Clearing Block}}\\
\hspace{1em}Resource constraint & $y_t = c_t$\\
\hspace{1em}Production function & $y_t = a_t + n_t$\\
\hspace{1em}Taylor rule & $r_t = \phi_\pi \pi^p_t - \mathbb{E}_t \pi^p_{t+1} - m_t$\\
\bottomrule
\end{tabular}

}

\end{table}%

\textbf{Exogenous shocks.}

Three exogenous structural shocks drive the system: a neutral technology
shock (\(a_t\)), a monetary policy shock (\(m_t\)), and a transfer shock
(\(z_t\)). Each obeys an independent AR(1) process: \[
\begin{aligned}
a_t &= \rho_a a_{t-1} + e^a_t \\
m_t &= \rho_m m_{t-1} + e^m_t \\
z_t &= \rho_z z_{t-1} + e^z_t
\end{aligned}
\] where the persistence parameters satisfy \(0 \le \rho_i < 1\) for
\(i \in \{a,m,z\}\), and the innovations \(e^i_t\) are zero-mean, i.i.d.
shocks. In both the log-linearized and canonical systems, \(z_t\) enters
as an exogenous transfer disturbance.

\subsection{The Canonical
Representation}\label{the-canonical-representation}

Starting from the nonredundant system above, I derive the reduced
representation used in the main analysis. The reduction yields a core of
five dynamic equations plus the static consumption-gap mapping needed to
close the TANK IS curve. It separates the aggregate New Keynesian block
from the two distributional objects of interest. The static mapping
isolates the earnings-reallocation term; the wage-gap law separates an
inherited wage-gap state from expected equilibrium-wedge forcing.

The reduction has three steps. First, a static distributional accounting
block separates direct profit-transfer incidence from endogenous
labor-income reallocation. Second, the wage block determines the current
wage gap from its own lag and the expected path of the net
distributional wedge. Finally, the distributional block is combined with
the TANK IS curve, the price Phillips curve, the aggregate real wage
equation, and the Taylor rule. Appendix A.1 gives the algebraic
derivation.

Table~\ref{tbl-derived-coefficients} defines the composite coefficients
used in the reduction, including \(\Theta_\sigma\) and \(\Theta_w\);
these coefficients scale current adjustment in the wage-gap and
aggregate-wage equations. Persistence depends on the corresponding
characteristic roots rather than on these coefficients alone.

\begin{table}

\caption{\label{tbl-derived-coefficients}Composite coefficients}

\centering{

\begin{tabular}[t]{ll}
\toprule
Symbol & Meaning\\
\midrule
$\kappa_w$ & Wage Phillips curve slope, $\psi_w/\eta_w$\\
$\kappa_p$ & Price Phillips curve slope, $\psi_p/\eta_p$\\
$\kappa_x$ & Wage sensitivity to the output gap, $\kappa_w(\gamma+\varphi)$\\
$\kappa_a$ & Composite wage-equation coefficient, $\kappa_p+\kappa_w$\\
$\Theta_\sigma$ & Wage-gap adjustment coefficient, $1+\beta+\kappa_w[(1-\gamma)+\psi_w(\gamma+\varphi)]$\\
\addlinespace
$\Theta_w$ & Aggregate wage adjustment factor, $1+\beta+\kappa_a$\\
\bottomrule
\end{tabular}

}

\end{table}%

\textbf{The earnings-reallocation term.}

Markup fluctuations cause aggregate profits to deviate from zero, so
\(d_t = a_t - w_t \neq 0\). I define the \emph{net distributional
wedge}, \(\omega_t\), as the component of the profit share that
government transfers do not neutralize: \[ \omega_t = d_t - z_t \] with
\(d_t - z_t \equiv a_t - w_t - z_t\).

The Type H budget constraint, together with \(c_t=w_t+n_t+d_t\) as
implied by the resource, production, and profit-share identities in
Table~\ref{tbl-full-system}, gives \[
\begin{aligned}
c_t^H-c_t
&=(w_t^H-w_t)+(n_t^H-n_t)-\omega_t\\
&=(1-\psi_w)(w_t^H-w_t)-\omega_t,
\end{aligned}
\] where the second equality follows from type-specific labor demand.
Since \(c_t^H-c_t=-(1-\lambda)\sigma_t^c\) and
\(w_t^H-w_t=-(1-\lambda)\sigma_t^w\), this gives the decomposition used
below (Appendix A.1):
\begin{equation}\protect\phantomsection\label{eq-static-dispersion}{
\sigma_t^c
=
\underbrace{\frac{\omega_t}{1-\lambda}}_{\text{direct profit-transfer incidence}}
+
\underbrace{(1-\psi_w)\sigma_t^w}_{\text{earnings-reallocation term}}.
}\end{equation} The first term is the consumption gap that would arise
if unequal profit exposure passed directly through to households. The
second term is the cross-type labor-income gap, since
\(\sigma_t^{w+n}=(1-\psi_w)\sigma_t^w\). When one type's relative wage
rises, its relative employment falls according to
\(\sigma_t^n=-\psi_w\sigma_t^w\). Because \(\psi_w>1\), the employment
response dominates the wage component of relative earnings. In the
benchmark monetary experiment below, this earnings movement opposes
direct profit-transfer incidence and therefore acts as a buffer that
compresses the consumption gap. The buffer label describes this
accounting role; it does not by itself imply a welfare ranking.

This reduced-form transfer wedge is closely related to Bilbiie (2020)
and Bilbiie (2025). Type-specific labor demand prevents the
profit-transfer term from passing one for one into the consumption gap.
Nominal wage adjustment then determines whether the
earnings-reallocation term remains static or persists through the
dynamic wage-gap block.

The flexible type-specific wage limit produces a static version of this
offset. As \(\eta_w\to0\) with finite \(\psi_w\), the wage-gap block
becomes static and satisfies \[
\sigma_t^w
\to
\frac{\gamma}{(1-\lambda)[1+\gamma(\psi_w-1)+\psi_w\varphi]}\omega_t,
\qquad
\sigma_t^c
\to
A_0(\psi_w)\frac{\omega_t}{1-\lambda},
\] where \[A_0(\psi_w)
\equiv
\frac{1+\psi_w\varphi}{1+\gamma(\psi_w-1)+\psi_w\varphi}
\in(0,1).\] The earnings term therefore equals
\(-[1-A_0(\psi_w)]\omega_t/(1-\lambda)\). It offsets a strictly positive
share of direct incidence for every nonzero \(\omega_t\), regardless of
its sign or the shock that generates it.

In the joint flexible-wage and perfect-substitutability limit
\(\psi_w\to\infty\) with \(\eta_w/\psi_w\to0\), the cross-type wage gap
vanishes, but the labor-reallocation effect converges to
\(-\frac{\gamma}{\varphi+\gamma}\frac{\omega_t}{1-\lambda}\). The
consumption gap then converges to
\begin{equation}\protect\phantomsection\label{eq-flex-wage-limit}{ \sigma_t^c \to \frac{\varphi}{\varphi+\gamma} \frac{\omega_t}{1-\lambda} }\end{equation}
Thus, flexible type-specific wages attenuate rather than eliminate the
consumption gap unless transfers neutralize the net wedge itself. The
finite-\(\psi_w\) flexible-wage limit is the Sticky-Price-Only TANK
restriction used in Appendix C.2. The joint flexible-wage and
perfect-substitutability limit isolates the same mechanism when the
labor types become arbitrarily substitutable.

\textbf{The inherited wage-gap state.}

With type-specific nominal wage rigidity, the wage gap is no longer
determined statically by contemporaneous fundamentals. Subtracting the
aggregate wage Phillips curve from its Type H counterpart and using the
own-lag wage-inflation identities and the markup definitions gives \[
\pi_t^H-\pi_t^w
=-(1-\lambda)(\sigma_t^w-\sigma_{t-1}^w),
\qquad
\mu_t^H-\mu_t^w
=-(1-\lambda)[(1-\gamma)+\psi_w(\gamma+\varphi)]\sigma_t^w+\gamma\omega_t.
\] Substituting these expressions into the difference between the two
wage Phillips curves yields
\begin{equation}\protect\phantomsection\label{eq-1}{ \sigma_t^w = \frac{1}{\Theta_\sigma}\left(\sigma_{t-1}^w + \beta \mathbb{E}_t\sigma_{t+1}^w + \dfrac{\gamma\kappa_w}{1-\lambda}\omega_t\right) }\end{equation}
Here, \(\kappa_w \equiv \psi_w/\eta_w\) is the slope of the wage
Phillips curve, and
\(\Theta_\sigma \equiv 1+\beta+\kappa_w[(1-\gamma)+\psi_w(\gamma+\varphi)]\)
is the wage-gap adjustment coefficient. The predetermined variable at
date \(t\) is \(\sigma_{t-1}^w\); the current wage gap \(\sigma_t^w\) is
jointly determined by that inherited state and the expected wedge path
from date \(t\) onward.

The lagged and expected terms in (\ref{eq-1}) arise from different
features of the maintained own-lag specification. Defining each type's
wage inflation relative to its own previous wage gives rise to the
inherited-state term \(\sigma_{t-1}^w\). Dynamic wage optimization
generates \(\beta\mathbb{E}_t\sigma_{t+1}^w\) because the current wage
also affects next period's adjustment cost. Forward iteration therefore
makes the current wage gap depend on both its inherited state and the
expected path of the distributional wedge.

Section~\ref{sec-perfect-alignment-main} derives a stable solution that
separates these two components within the same baseline model; the
quantitative analysis then measures their contributions separately. As
\(\eta_w \to 0\) with finite \(\psi_w\), the wage-gap law collapses back
to a static mapping of the contemporaneous wedge.

\textbf{Macroeconomic blocks and composite parameters.}

The macroeconomic blocks mirror standard New Keynesian equations but add
distributional feedback. Let \(x_t=y_t-y_t^f\), where
\(y_t^f\equiv(1+\varphi)a_t/(\gamma+\varphi)\) is aggregate output when
both prices and wages are flexible. This expression remains valid under
transfers in the symmetric first-order economy: zero price and wage
markup deviations imply \(w_t^f=a_t=\gamma y_t^f+\varphi(y_t^f-a_t)\).
Transfers change the type-specific flexible allocation and its
consumption gap without changing this aggregate natural-output
expression. Thus \(x_t\) is measured relative to the jointly flexible
price-and-wage allocation; it is not a gap relative to an allocation
that retains sticky wages. The linear technology is an expositional
benchmark. The online appendix shows that the same five-equation
canonical structure obtains under \(Y_t=A_tN_t^{1-\alpha}\) once the
flexible-allocation objects and composite coefficients are redefined. In
that extension, a steady-state lump-sum transfer equalizes disposable
income across the two household types. Production curvature therefore
changes the coefficients, not the mechanism.

To distinguish the technology component from the full natural rate,
define \begin{equation}\protect\phantomsection\label{eq-natural-rates}{
\begin{aligned}
r_t^R&\equiv\gamma\mathbb{E}_t\Delta y^f_{t+1}
=\gamma\frac{1+\varphi}{\gamma+\varphi}\mathbb{E}_t\Delta a_{t+1},\\
r_t^n&\equiv r_t^R+\gamma\lambda\mathbb{E}_t\Delta\sigma_{t+1}^{c,f},
\end{aligned}
}\end{equation} where \(\sigma_t^{c,f}\) is the consumption gap in the
allocation with both prices and wages flexible under the same
wage-setting institution and fiscal rule. The full TANK natural rate
\(r_t^n\) follows from the saver Euler equation evaluated at that
allocation. Transfers can change it through expected flexible-allocation
consumption-gap growth even though they leave aggregate natural output
unchanged.

The aggregate demand block yields the TANK IS curve:
\begin{equation}\protect\phantomsection\label{eq-tank-is}{ x_t = \mathbb{E}_t x_{t+1} - \frac{1}{\gamma}(r_t - r_t^R) - \lambda (\sigma^c_t - \mathbb{E}_t \sigma^c_{t+1}) }\end{equation}
Equivalently, subtracting the flexible-allocation Euler equation gives
\begin{equation}\protect\phantomsection\label{eq-tank-is-natural}{
\begin{aligned}
x_t={}&\mathbb{E}_t x_{t+1}-\frac{1}{\gamma}(r_t-r_t^n)\\
&-\lambda\left[(\sigma_t^c-\sigma_t^{c,f})
-\mathbb{E}_t(\sigma_{t+1}^c-\sigma_{t+1}^{c,f})\right].
\end{aligned}
}\end{equation} This form measures both aggregate output and consumption
dispersion relative to the jointly flexible allocation. The canonical
system below uses the equivalent \(r_t^R\) representation, retaining the
actual consumption gap. Since the flexible consumption gap can differ
across wage-setting institutions, their full natural rates need not
coincide. The term \(\lambda\mathbb{E}_t\Delta\sigma^c_{t+1}\) in
(\ref{eq-tank-is}) is the heterogeneity channel through which expected
changes in consumption dispersion affect current aggregate demand.

The macroeconomic supply block consists of the price Phillips curve and
a unified aggregate real-wage equation: \[\begin{aligned}
\pi^p_t &= \kappa_p(w_t - a_t) + \beta \mathbb{E}_t \pi^p_{t+1}, \\
w_t &= \frac{1}{\Theta_w}\left(w_{t-1} + \beta \mathbb{E}_t w_{t+1} + \kappa_x x_t + \kappa_a a_t\right).
\end{aligned}\] The composite parameters combine the effects of the two
nominal rigidities. The price Phillips curve slope is
\(\kappa_p \equiv \psi_p/\eta_p\), so greater price rigidity flattens
the curve. The wage-block slope \(\kappa_w\) determines the output-gap
loading in real wages through
\(\kappa_x \equiv \kappa_w(\gamma+\varphi)\). When wages are highly
rigid (\(\eta_w \to \infty\)), \(\kappa_w \to 0\) and
\(\kappa_x \to 0\). The real wage then becomes insensitive to the output
gap, and the standard labor-market adjustment mechanism is suppressed.
The remaining terms are \(\kappa_a \equiv \kappa_p + \kappa_w\), the
composite technology-loading coefficient, and
\(\Theta_w \equiv 1+\beta+\kappa_a\), the aggregate wage adjustment
factor.

The wage-gap law and the aggregate real-wage equation have parallel
second-order expectational forms governed by \(\Theta_\sigma\) and
\(\Theta_w\). Their difference is
\(\Theta_\sigma-\Theta_w=\kappa_w[\psi_w(\gamma+\varphi)-\gamma]-\kappa_p\).
Thus \(\Theta_\sigma>\Theta_w\) if and only if
\(\kappa_w[\psi_w(\gamma+\varphi)-\gamma]>\kappa_p\), which holds at the
benchmark calibration. Holding the variables on the respective
right-hand sides fixed, the larger \(\Theta_\sigma\) dampens the current
wage gap more strongly than \(\Theta_w\) dampens the aggregate real
wage. Persistence is governed by the stable characteristic root rather
than by \(\Theta\) alone. At the benchmark calibration
(Table~\ref{tbl-wage-roots}), the smaller stable root of the wage-gap
equation implies that the inherited wage gap decays faster than the
aggregate real wage.

Monetary policy follows a standard Taylor rule. The coefficient
\(\phi_\pi > 1\)\footnote{ The condition \(\phi_\pi > 1\) ensures the
  Taylor principle, which is necessary for local determinacy in the RANK
  limit. In the heterogeneous-agent economy, local determinacy of the
  full canonical system is verified numerically in Appendix C.1.} is the
central bank's inflation response, and \(m_t\) is an exogenous monetary
policy shock:
\[ r_t = \phi_\pi \pi^p_t - \mathbb{E}_t \pi^p_{t+1} - m_t \]

\textbf{The canonical system.}

Table~\ref{tbl-canonical} summarizes the five dynamic equations and the
auxiliary static mapping that closes the system. Keeping the
distributional equations separate from the aggregate equations clarifies
where heterogeneity enters transmission. Here \(r_t^R\) is the
technology/RANK real-rate component defined above,
\(\omega_t=d_t-z_t=a_t-w_t-z_t\) is the net distributional wedge, and
\(\sigma_t^c\) and \(\sigma_t^w\) denote the signed cross-type
consumption and real-wage gaps. Appendix A.1 provides the detailed
elimination steps and recovery identities.

\begin{table}

\caption{\label{tbl-canonical}Canonical system}

\centering{

\begin{tabular}[t]{ll}
\toprule
Name & Equation\\
\midrule
TANK IS curve & $x_t = \mathbb{E}_t x_{t+1} - \frac{1}{\gamma}(r_t - r_t^R) - \lambda (\sigma_t^c - \mathbb{E}_t \sigma_{t+1}^c)$\\
Price Phillips curve & $\pi_t^p = \beta \mathbb{E}_t \pi_{t+1}^p +\kappa_p(w_t-a_t)$\\
Taylor rule & $r_t = \phi_\pi \pi_t^p - \mathbb{E}_t \pi_{t+1}^p - m_t$\\
Aggregate real wage equation & $w_t = \frac{1}{\Theta_w}\left(w_{t-1} + \beta \mathbb{E}_t w_{t+1} + \kappa_x x_t + \kappa_a a_t\right)$\\
Wage-gap law of motion & $\sigma_t^w = \frac{1}{\Theta_\sigma}\left(\sigma_{t-1}^w + \beta \mathbb{E}_t\sigma_{t+1}^w + \dfrac{\gamma\kappa_w}{1-\lambda}\omega_t\right)$\\
\addlinespace
Auxiliary static consumption-gap mapping & $\sigma_t^c = (1-\psi_w)\sigma_t^w + \dfrac{\omega_t}{1-\lambda}$\\
\bottomrule
\end{tabular}

}

\end{table}%

The canonical system nests the limiting cases used below. When
\(\lambda=0\), the distributional block decouples from aggregate
dynamics and the model reduces to a RANK economy with dual nominal
rigidities:
\begin{equation}\protect\phantomsection\label{eq-rank-is}{ r_t = r_t^R + \gamma \mathbb{E}_t\Delta x_{t+1} }\end{equation}
In this limit, \(r_t^R\) is the RANK natural real rate because the
consumption-gap term vanishes.

With transfers set to zero, the flexible-wage limit recovers the
IS-slope inversion condition analyzed in Miyazaki (2026), Section 4.2.
This slope condition is distinct from local determinacy, which is
assessed separately for the full own-lag system in Appendix C.1.

In the further joint flexible-wage and perfect-substitutability limit
\(\psi_w\to\infty\) (with \(\eta_w/\psi_w\to0\)), the cross-type wage
gap vanishes (\(\sigma_t^w\to0\)). As (\ref{eq-flex-wage-limit}) shows,
the consumption gap need not vanish. Substituting this limiting
consumption-gap expression into the aggregate demand block gives the
corresponding IS curve:
\[ x_t = \mathbb{E}_t x_{t+1} - \frac{1}{\gamma}(r_t-r_t^R) + \frac{\lambda\varphi}{(\varphi+\gamma)(1-\lambda)}\mathbb{E}_t\Delta\omega_{t+1} \]
Hence, the limiting economy remains a sticky-price-only TANK model
rather than a RANK model unless the net wedge itself is fully
neutralized (\(\omega_t=0\)). Eliminating the wage gap is insufficient
if the profit-transfer wedge continues to generate a consumption
gap.\footnote{ This limiting result differs from equivalence results
  that compare sticky-price and sticky-wage polar cases in
  representative-agent environments, such as Bilbiie and Trabandt
  (2025). With \(\lambda>0\), the condition for collapse to RANK is not
  merely \(\sigma_t^w=0\), but \(\omega_t=0\).}

\subsection{A Reduced-Form Common-Wage
Contrast}\label{a-reduced-form-common-wage-contrast}

To define the reduced-form contrast, impose a common wage across
household types, \(w_t^H=w_t^S=w_t\). Under CES labor demand, equal
wages imply equal employment, \(n_t^H=n_t^S=n_t\); this equal-employment
outcome therefore follows from the wage closure rather than being
imposed independently. The implemented closure removes the Type H
wage-setting block and retains the aggregate sticky-wage equation. The
comparison eliminates cross-type wage and employment reallocation while
preserving aggregate wage adjustment. It is neither a fully derived
alternative wage-setting institution nor a nested restriction obtained
by varying \(\eta_w\).

This setup is related to homogeneous-labor-supply arrangements in
sticky-wage HANK models with union wage setting (Gerke et al. 2024) and
to the common-wage sticky-wage TANK model of Diz et al. (2023). Bojovic
and Ellis (2026) retain a common efficiency-unit wage and equal hours
across financial types. They enrich hand-to-mouth disposable income by
introducing productivity differences, an allocation of firm profits to
hand-to-mouth households, and debt-service exposure. Neither TANK model
contains the financial-type relative-wage labor-demand margin studied
here.

The closure used here preserves the aggregate sticky-wage equation and
the profit-transfer wedge but omits the additional productivity,
profit-allocation, and balance-sheet blocks of Bojovic and Ellis (2026).
It therefore does not replicate either model. The static consumption-gap
mapping reduces to
\begin{equation}\protect\phantomsection\label{eq-common-wage-static}{
\sigma_t^c = \frac{\omega_t}{1-\lambda}, \qquad \omega_t = a_t - w_t - z_t.
}\end{equation}

Under this closure, the canonical system no longer contains an
independent wage-gap law. The first four dynamic rows of
Table~\ref{tbl-canonical} remain unchanged. The fifth row (the wage-gap
law) is removed, and the sixth row reduces to
(\ref{eq-common-wage-static}). Equivalently, substituting
(\ref{eq-common-wage-static}) into the TANK IS curve in
Table~\ref{tbl-canonical} yields
\begin{equation}\protect\phantomsection\label{eq-common-wage-is}{
x_t = \mathbb{E}_t x_{t+1} - \frac{1}{\gamma}(r_t-r_t^R) - \frac{\lambda}{1-\lambda}\left(\omega_t-\mathbb{E}_t\omega_{t+1}\right).
}\end{equation}

The economy remains TANK as long as \(\omega_t \neq 0\), but both the
earnings-reallocation term and the independent wage-gap state disappear.
The contrast makes the distinction explicit: persistent aggregate wages
remain present, whereas cross-type earnings reallocation and the
inherited wage-gap state are absent. Quantitative differences between
the two closures reflect both this missing margin and the accompanying
general-equilibrium adjustment. Appendix A.3 provides the algebraic
reduction.

\section{The Distributional Buffer}\label{the-distributional-buffer}

Type-specific labor demand generates an earnings-reallocation term that
can offset the direct incidence of profits and transfers. Comparing
perfect alignment with mixed labor segments clarifies how substitution
and financial-type composition determine the sign and size of this term.

\subsection{Distributional Incidence under Perfect
Alignment}\label{sec-perfect-alignment-main}

In the benchmark, financial types coincide with labor-market segments,
and households set their variety wages subject to own-lag adjustment
costs. A single CES labor aggregator makes \(\psi_w\) govern both
individual wage demand and substitution across financial types. The same
elasticity therefore determines the unsubsidized wage markup and the
earnings response to relative wages.

Proposition~\ref{prp-stable-wage-gap} gives the conditional solution for
the wage gap, and Proposition~\ref{prp-impact-attenuation} states a
necessary and sufficient condition for a strict partial offset when the
inherited wage gap is zero. Corollary~\ref{cor-ar1-wedge} shows that an
AR(1)-type conditional wedge path is sufficient. Own-lag wage adjustment
also introduces a predetermined wage-gap state. This state is distinct
from the expected path of the distributional wedge and need not be the
dominant source of equilibrium persistence. The aggregate wage equation
determines the wedge jointly with the rest of the equilibrium system, so
the condition can be evaluated without assuming a scalar autoregression
for the wedge.

\textbf{Conditional wage-gap dynamics.}

For any given path of the net distributional wedge, the wage block
determines the extent to which labor-income reallocation offsets direct
incidence. Write the wage-gap equation (\ref{eq-1}) as \[
\beta \mathbb{E}_t\sigma_{t+1}^w
-\Theta_\sigma\sigma_t^w
+\sigma_{t-1}^w
=
-\frac{\gamma\kappa_w}{1-\lambda}\omega_t.
\]

\begin{proposition}[Stable Wage-Gap
Dynamics]\protect\hypertarget{prp-stable-wage-gap}{}\label{prp-stable-wage-gap}

Assume that \(0<\beta<1\), \(0<\lambda<1\), \(\psi_w>1\), \(\gamma>0\),
and \(\varphi>0\). Suppose that \(0<\eta_w<\infty\) and that both the
full log-linearized economy and its canonical representation are well
defined. Given an inherited state \(\sigma_{t-1}^w\), the wage-gap
equation admits a unique non-explosive solution for every bounded
conditional path of \(\omega_t\) if \(\Theta_\sigma>1+\beta\). Let
\(\xi_1\in(0,1)\) and \(\xi_2>1\) denote the roots of \[
\beta\xi^2-\Theta_\sigma\xi+1=0.
\] The stable solution is
\begin{equation}\protect\phantomsection\label{eq-wage-gap-forward-solution}{
\sigma_t^w
=
\underbrace{\xi_1\sigma_{t-1}^w}_{\text{contribution of inherited wage-gap state}}
+
\underbrace{
\frac{\gamma\kappa_w}{(1-\lambda)\beta\xi_2}
\sum_{k=0}^{\infty}
\left(\frac{1}{\xi_2}\right)^k
\mathbb{E}_t\omega_{t+k}
}_{\text{expected equilibrium-wedge forcing}}.
}\end{equation}

\end{proposition}

The proof of Proposition~\ref{prp-stable-wage-gap} is in Appendix B.1.
The root condition holds under the maintained assumptions because \[
\Theta_\sigma-(1+\beta)
=
\kappa_w\bigl[(1-\gamma)+\psi_w(\gamma+\varphi)\bigr]
=
\kappa_w\bigl[1+\gamma(\psi_w-1)+\psi_w\varphi\bigr]
>0
\] under \(\psi_w>1\), \(\gamma>0\), \(\varphi>0\), and \(\kappa_w>0\).
Equation (\ref{eq-wage-gap-forward-solution}) yields an exact
conditional decomposition within the log-linearized model. It clarifies
the mechanism by separating the two components combined in an impulse
response. The first term carries the inherited wage gap into period
\(t\). The second depends on the entire expected equilibrium path of
\(\omega_t=a_t-w_t-z_t\). The equation neither determines that path nor,
by itself, guarantees a partial offset. Whether an offset occurs, and
how large it is, must be assessed from the full equilibrium solution.

\Needspace{4\baselineskip}

\textbf{The equilibrium wedge.}

Let \(\mathfrak f_t^w\equiv\kappa_xx_t+\kappa_aa_t\) denote the forcing
term in the aggregate real-wage equation. The same forward-iteration
argument used in Proposition~\ref{prp-stable-wage-gap} gives
\begin{equation}\protect\phantomsection\label{eq-equilibrium-wage-wedge}{
\begin{aligned}
w_t
&=\chi_1w_{t-1}
+\frac{1}{\beta\chi_2}
\sum_{k=0}^{\infty}\chi_2^{-k}\mathbb{E}_t\mathfrak f_{t+k}^w,\\
\omega_t
&=a_t-z_t-\chi_1w_{t-1}
-\frac{1}{\beta\chi_2}
\sum_{k=0}^{\infty}\chi_2^{-k}\mathbb{E}_t\mathfrak f_{t+k}^w,
\end{aligned}
}\end{equation} where \(\chi_1\in(0,1)\) and \(\chi_2>1\) solve
\(\beta\chi^2-\Theta_w\chi+1=0\). These roots exist because
\(\Theta_w=1+\beta+\kappa_a>1+\beta\). The solution is conditional on
the equilibrium output-gap path and the non-explosive boundary
condition; by itself, it does not establish determinacy of the full
economy. Appendix B.1 gives the derivation.

Equation (\ref{eq-equilibrium-wage-wedge}) makes the sources of wedge
dynamics explicit. The wedge reflects the inherited aggregate wage
state, external disturbances, and the expected path of output. Although
the aggregate wage equation is second order, its forcing term is
endogenous. Consequently, the model does not generally imply a geometric
conditional law for \(\omega_t\).

\begin{proposition}[Criterion for Equilibrium Impact
Attenuation]\protect\hypertarget{prp-impact-attenuation}{}\label{prp-impact-attenuation}

Under the assumptions of Proposition~\ref{prp-stable-wage-gap}, suppose
that \(\sigma_{t-1}^w=0\) and \(\omega_t\neq0\). When the following
discounted sum converges, define
\begin{equation}\protect\phantomsection\label{eq-wedge-present-value}{
\operatorname{PV}_t[\omega]
\equiv\sum_{k=0}^{\infty}\xi_2^{-k}\mathbb{E}_t\omega_{t+k},
\qquad
\mathcal M_t^\omega
\equiv\frac{\operatorname{PV}_t[\omega]}{\omega_t},
}\end{equation} The operator applies in the same way to any other scalar
path for which the sum converges. Let
\begin{equation}\protect\phantomsection\label{eq-equilibrium-offset-share}{
K_\omega\equiv(\psi_w-1)\gamma\kappa_w,
\qquad
\mathcal B_t^{\mathrm{GE}}
\equiv\frac{K_\omega}{\beta\xi_2}\mathcal M_t^\omega.
}\end{equation} Then
\begin{equation}\protect\phantomsection\label{eq-equilibrium-impact-decomposition}{
\begin{aligned}
(1-\psi_w)\sigma_t^w
&=-\mathcal B_t^{\mathrm{GE}}\frac{\omega_t}{1-\lambda},\\
\sigma_t^c
&=(1-\mathcal B_t^{\mathrm{GE}})\frac{\omega_t}{1-\lambda}.
\end{aligned}
}\end{equation} The earnings term opposes direct incidence, strictly
reduces the consumption gap in absolute value, and leaves the gap with
the same sign as direct incidence if and only if
\begin{equation}\protect\phantomsection\label{eq-equilibrium-offset-condition}{
0<\mathcal B_t^{\mathrm{GE}}<1
\quad\Longleftrightarrow\quad
0<\mathcal M_t^\omega<\frac{\beta\xi_2}{K_\omega}.
}\end{equation}

\end{proposition}

Proposition~\ref{prp-impact-attenuation} follows from substituting
(\ref{eq-wage-gap-forward-solution}) into (\ref{eq-static-dispersion});
Appendix B.2 provides the proof and an equivalent condition in terms of
the aggregate-wage forcing. The proposition characterizes when the
offset is partial; it does not imply that every equilibrium wedge path
satisfies the condition. In general, evaluating this condition requires
solving the full equilibrium, typically numerically.

Relative labor demand gives this condition its economic meaning. When
the discounted expected wedge has the same sign as the current wedge,
the wage gap moves in the same direction. Labor demand shifts away from
the type with the relatively higher wage; because \(\psi_w>1\), the
resulting employment loss outweighs that type's wage gain in relative
earnings. The lower bound in (\ref{eq-equilibrium-offset-condition})
therefore ensures that the two terms have opposite signs, while the
upper bound rules out complete or excessive offsetting. The condition
applies symmetrically to either sign of direct incidence.

\begin{corollary}[Impact Attenuation under an AR(1)-Type Wedge
Path]\protect\hypertarget{cor-ar1-wedge}{}\label{cor-ar1-wedge}

Under the assumptions of Proposition~\ref{prp-impact-attenuation},
including \(\sigma_{t-1}^w=0\) and \(\omega_t\neq0\), suppose further
that \[
\mathbb{E}_t\omega_{t+k}=\rho_\omega^k\omega_t,
\qquad k\geq0,\qquad 0\leq\rho_\omega<1.
\] Then \(\mathcal M_t^\omega=(1-\rho_\omega/\xi_2)^{-1}\) and
\begin{equation}\protect\phantomsection\label{eq-geometric-offset-share}{
\mathcal B_t^{\mathrm{GE}}
=\mathcal B(\rho_\omega)
\equiv\frac{K_\omega}{\beta(\xi_2-\rho_\omega)}
\in(0,1).
}\end{equation} Hence the condition in
Proposition~\ref{prp-impact-attenuation} is satisfied, and earnings
reallocation provides a strict partial offset to direct incidence
without reversing its sign.

\end{corollary}

Appendix B.2 proves Corollary~\ref{cor-ar1-wedge}, including the strict
upper bound. The stated conditional path arises when \(\omega_t\) itself
follows a zero-mean AR(1) process with coefficient \(\rho_\omega\) and
innovations whose conditional future means are zero. Here
\(\rho_\omega\) denotes the wedge's conditional decay rate, not an
exogenous structural-shock parameter. AR(1) processes for the structural
disturbances alone do not imply that the endogenous wedge satisfies this
restriction.\footnote{For a related static attenuation result under
  aggregate-lag wage adjustment, see Miyazaki (2026, Proposition 1).
  That result does not require a geometric wedge path, whereas
  Corollary~\ref{cor-ar1-wedge} establishes attenuation under own-lag
  adjustment subject to the stated path and initial-state restrictions.}

\textbf{Evaluation in the full equilibrium.}

Write a stable solution in the form \[
\mathbf s_t=\mathbf T_{\mathrm{eq}}\mathbf s_{t-1}
+\mathbf J_{\mathrm{sh}}\boldsymbol\varepsilon_t,
\qquad
\omega_t=\mathbf v_\omega'\mathbf s_t,
\] where \(\boldsymbol\varepsilon_t=(e_t^a,e_t^m,e_t^z)'\) collects the
structural innovations, written componentwise as
\(\varepsilon_{i,t}\equiv e_t^i\). The innovations have zero conditional
future means, and \(\operatorname{spr}(\mathbf T_{\mathrm{eq}})<1\). The
multiplier has the following finite-dimensional expression:
\begin{equation}\protect\phantomsection\label{eq-equilibrium-resolvent}{
\mathcal M_t^\omega
=\frac{\mathbf v_\omega'
(I-\mathbf T_{\mathrm{eq}}/\xi_2)^{-1}\mathbf s_t}
{\mathbf v_\omega'\mathbf s_t}.
}\end{equation} In the baseline canonical system,
\(\mathbf s_t=(z_t,m_t,w_t,\sigma_t^w,a_t)'\) and
\(\mathbf v_\omega=(-1,0,-1,0,1)'\). Starting from the steady state, a
single innovation of type \(i\in\{a,m,z\}\) gives
\(\mathbf s_0=\mathbf b_i\varepsilon_{i,0}\), where \(\mathbf b_i\) is
the corresponding column of \(\mathbf J_{\mathrm{sh}}\). Its magnitude
cancels from (\ref{eq-equilibrium-resolvent}). The impact condition can
therefore be checked for each shock using the full equilibrium solution
without truncating an impulse response.

Stability ensures that the discounted sum converges, but it does not
impose the sign or upper bound required by
Proposition~\ref{prp-impact-attenuation} in
(\ref{eq-equilibrium-offset-condition}). At the benchmark, the condition
holds separately for each of the three structural innovations, as
established in Section 4 and Appendix B.3. Because the impact ratios
differ across shocks, the condition need not hold for arbitrary
simultaneous mixtures, even when the inherited states are zero. With a
nonzero inherited wage gap, the exact decomposition instead becomes
\begin{equation}\protect\phantomsection\label{eq-equilibrium-inherited-decomposition}{
\sigma_t^c
=(1-\psi_w)\xi_1\sigma_{t-1}^w
+(1-\mathcal B_t^{\mathrm{GE}})\frac{\omega_t}{1-\lambda}.
}\end{equation} Accordingly, later-horizon accounting shares include the
inherited contribution and are evaluated separately.

The term \emph{buffer} refers only to this incidence accounting. It does
not imply that type-specific wages improve welfare. The model tracks a
signed cross-type consumption gap and changes in labor allocation; it
does not aggregate the utility losses associated with consumption
dispersion, employment movements, or wage-adjustment costs.

\textbf{The inherited state.}

Own-lag wage adjustment introduces the first term in
(\ref{eq-wage-gap-forward-solution}). At date \(t\), the predetermined
object is \(\sigma_{t-1}^w\); the current wage gap \(\sigma_t^w\)
combines this inherited state with the expected wedge path. Holding the
expected path fixed, \[
\frac{\partial\sigma_t^w}{\partial\sigma_{t-1}^w}
=
\xi_1.
\] The corresponding conditional half-life is \[
\frac{\log(1/2)}{\log(\xi_1)}.
\] This half-life captures the persistence generated by the own-lag
state alone. By contrast, the half-life of an impulse response of
\(\sigma_t^w\) to a shock also reflects the shock's persistence and the
endogenous path of \(\omega_t\).

\subsection{Separate Labor Elasticities and Partial
Alignment}\label{sec-segmentation-main}

The extension allows each labor segment to contain both financial types
and assigns wage setting to unions under a specified delegation mandate.
The segments have population shares \(\lambda\) and \(1-\lambda\). Their
fixed fractions of Type H members are
\begin{equation}\protect\phantomsection\label{eq-main-segment-composition}{
q_1=\lambda+\delta(1-\lambda),\qquad
q_2=\lambda(1-\delta),\qquad 0\le\delta\le1.
}\end{equation} The overall Type H share remains \(\lambda\), and
\(\delta=q_1-q_2\) measures the difference in the two segments' Type H
shares. This alignment parameter is distinct from production curvature
\(\alpha\). Perfect alignment is \(\delta=1\); identical financial-type
composition is \(\delta=0\).

A continuum of infinitesimal unions sets variety wages using
marginal-utility weights delegated by the two financial-type income
pools. Within each variety, members work equal hours and receive equal
labor earnings; income is pooled within financial type, with no asset
trade by the Type H pool. The online appendix states the union
objective, subsidies, income allocation, and nonlinear budget
constraints. This is an explicit extension of the wage-setting
institution, not a claim that partial mixing preserves the original
individual-household optimization problem unchanged.

A nested CES labor aggregator separates the two elasticities tied
together in the benchmark. Let \(\varepsilon_w>1\) govern demand for
individual wage-setting varieties within each labor segment and let
\(\nu>0\) govern substitution between the two segments. Wage subsidies
offset the within-segment markup, and the wage Phillips-curve slope is
\(\kappa_w=\varepsilon_w/\eta_w\). Between-segment labor demand instead
satisfies \(n_{2t}-n_{1t}=-\nu v_t\), where \(v_t=w_{2t}-w_{1t}\) is the
segment wage gap.

The financial-type mean wage and employment gaps are \(\delta v_t\) and
\(-\delta\nu v_t\). Their consumption gap is therefore
\begin{equation}\protect\phantomsection\label{eq-main-segment-incidence}{
\sigma_t^c=\frac{\omega_t}{1-\lambda}+\delta(1-\nu)v_t.
}\end{equation} The segment wage-gap equation becomes
\begin{equation}\protect\phantomsection\label{eq-main-segment-dynamics}{
\beta \mathbb{E}_tv_{t+1}-\Theta_v v_t+v_{t-1}
=-\frac{\gamma\kappa_w\delta}{1-\lambda}\omega_t,
}\end{equation} where
\(\Theta_v=1+\beta+\kappa_w[1+\nu\varphi+\gamma\delta^2(\nu-1)]\). The
aggregate equations are unchanged. At \(\delta=1\) and
\(\nu=\varepsilon_w=\psi_w\), the first-order system equals the
benchmark. At \(\delta=0\), financial-type earnings are equal and, from
a zero inherited segment gap, \(v_t=0\); the common-wage first-order
system follows. This is a structurally derived common-wage endpoint
within the delegated-union and within-type income-pooling institution.
Segment membership, alignment, pooling, and delegation are maintained
assumptions rather than endogenous institutional choices. The original
common-wage comparison remains a reduced-form change of wage-setting
block.

\begin{proposition}[The Sign of Segment-Earnings
Incidence]\protect\hypertarget{prp-segment-earnings-incidence}{}\label{prp-segment-earnings-incidence}

Assume \(0<\lambda<1\), \(0<\beta<1\), \(\gamma,\varphi,\eta_w>0\),
\(\varepsilon_w>1\), and
\begin{equation}\protect\phantomsection\label{eq-main-segment-root-condition}{
1+\nu\varphi+\gamma\delta^2(\nu-1)>0.
}\end{equation} Let \(\xi_{v,1}\in(0,1)\) and \(\xi_{v,2}>1\) solve
\(\beta\xi^2-\Theta_v\xi+1=0\). With \(v_{t-1}=0\) and \(\omega_t\ne0\),
the offset share is
\begin{equation}\protect\phantomsection\label{eq-main-segment-offset}{
\mathcal B_t^v=
\frac{\gamma\delta^2\kappa_w(\nu-1)}{\beta\xi_{v,2}}
\frac{\sum_{k\ge0}\xi_{v,2}^{-k}E_t\omega_{t+k}}{\omega_t}.
}\end{equation} If the discounted expected wedge has the sign of the
current wedge and \(\delta>0\), the earnings term offsets direct
incidence when \(\nu>1\) and amplifies it when \(\nu<1\). It is zero at
\(\nu=1\) or \(\delta=0\). Strict partial offset additionally requires
\(\mathcal B_t^v<1\). For a geometric conditional wedge path with
persistence in \([0,1)\) and \(\nu>1\), this upper bound holds; within
this geometric-path family, the share increases with alignment,
between-segment substitution, and wedge persistence, and decreases with
the wage-adjustment cost, holding the other parameters fixed.

\end{proposition}

The proof applies the same forward solution as
Proposition~\ref{prp-stable-wage-gap} and is given in the online
appendix. For \(\delta=1\), the root condition reduces to
\(1-\gamma+\nu(\gamma+\varphi)>0\); allowing \(\nu<1\) therefore
requires checking it rather than importing the single-CES root argument.
The comparative statics hold for a fixed conditional wedge path.
Re-solving general equilibrium also changes that path, so the numerical
experiments separately assess equilibrium incidence. These sign results
concern a zero inherited segment wage gap; nonzero inherited wages
contribute an additional term.

\section{Quantitative Analysis}\label{quantitative-analysis}

The quantitative analysis evaluates the partial-offset condition in the
benchmark equilibrium, traces the earnings-reallocation mechanism along
the resulting path, and compares consumption and output responses across
wage-setting closures. The within-path accounting holds the
type-specific-wage equilibrium fixed; the cross-closure comparison
re-solves the equilibrium under the reduced-form common-wage
restriction. A subsequent decomposition separates inherited-state
dynamics from expected equilibrium-wedge forcing to explain the
persistence of the wage gap.

\subsection{Quantitative Design and
Calibration}\label{quantitative-design-and-calibration}

The exercises use an illustrative quarterly New Keynesian calibration.
The original single-CES model is the \emph{nested perfect-alignment
benchmark}: in the extension, it sets \(\delta=1\) and
\(\nu=\varepsilon_w=\psi_w=10\). Unless stated otherwise, the benchmark
monetary magnitudes refer to this restriction. I neither estimate the
parameters nor match cross-type wage, employment, income, or liquidity
moments, so the reported levels are model benchmarks rather than
empirical estimates. The preference block sets \(\beta=0.99\),
\(\gamma=1\), and \(\varphi=1\). The benchmark hand-to-mouth share is
\(\lambda=0.25\), consistent with the analytical TANK literature
(Bilbiie 2008; Debortoli and Galí 2025). The benchmark sets the
Rotemberg price-adjustment cost to \(\eta_p=50\) and the common
wage-adjustment cost to \(\eta_w^H=\eta_w^S=\eta_w=50\). All sensitivity
exercises for \(\eta_w\) maintain \(\eta_w^H=\eta_w^S\) and therefore do
not cover asymmetric wage rigidity. The sticky-wage TANK literature
motivates including both nominal rigidities (Colciago 2011; Ida and
Okano 2024). The values used here are illustrative reduced-form
adjustment costs rather than Calvo durations, and the corresponding
slopes depend on the model's normalization and curvature terms.

The substitution elasticities \(\psi_p=10\) and \(\psi_w=10\) imply
desired markups of about 11 percent in the absence of subsidies; the
adopted subsidy rates make the steady-state gross markups equal to one.
In the single-CES benchmark, \(\psi_w\) governs both individual wage
demand and cross-type substitution, so a markup-based choice does not
separately discipline the reallocation elasticity. The nested extension
holds the individual wage-setting elasticity fixed while varying
substitution between segments. In particular, the markup interpretation
of \(\varepsilon_w=10\) supplies no independent empirical justification
for \(\nu=10\) or \(\delta=1\). These two parameters are not estimated
here. Table~\ref{tbl-segmentation-incidence} therefore also reports
partial alignment with lower segment substitution. The AR(1)
coefficients are set to \(0.8\), so the exogenous disturbances are
persistent but mean reverting.

\begin{table}[t]

\caption{\label{tbl-calibration}Nested perfect-alignment calibration}

\centering{

\centering
\begin{threeparttable}
\begin{tabular}[t]{>{\centering\arraybackslash}p{0.22\linewidth}>{\centering\arraybackslash}p{0.22\linewidth}}
\toprule
Symbol & Value\\
\midrule
$\beta$ & $0.99$\\
$\gamma$ & $1.0$\\
$\varphi$ & $1.0$\\
$\lambda$ & $0.25$\\
$\psi_p$ & $10.0$\\
\addlinespace
$\eta_p$ & $50.0$\\
$\psi_w$ & $10.0$\\
$\eta_w$ & $50.0$\\
$\phi_\pi$ & $1.5$\\
$\rho_a$ & $0.8$\\
\addlinespace
$\rho_m$ & $0.8$\\
$\rho_z$ & $0.8$\\
\bottomrule
\end{tabular}
\begin{tablenotes}
\item \textit{Notes:} Persistence parameters denote AR(1) coefficients.
\end{tablenotes}
\end{threeparttable}

}

\end{table}%

\Needspace{0.43\textheight}

An empirical application would require separate information on
individual wage demand, substitution between segments, and the
distribution of financial types across them.
Table~\ref{tbl-parameter-counterparts} maps these parameters to
potential observable moments in the extended model.

\begin{longtable}[]{@{}
  >{\raggedright\arraybackslash}p{(\linewidth - 2\tabcolsep) * \real{0.1400}}
  >{\raggedright\arraybackslash}p{(\linewidth - 2\tabcolsep) * \real{0.8600}}@{}}
\caption{Empirical counterparts of labor-market and financial
parameters}\label{tbl-parameter-counterparts}\tabularnewline
\toprule\noalign{}
\begin{minipage}[b]{\linewidth}\raggedright
Parameter
\end{minipage} & \begin{minipage}[b]{\linewidth}\raggedright
Potential empirical counterpart
\end{minipage} \\
\midrule\noalign{}
\endfirsthead
\toprule\noalign{}
\begin{minipage}[b]{\linewidth}\raggedright
Parameter
\end{minipage} & \begin{minipage}[b]{\linewidth}\raggedright
Potential empirical counterpart
\end{minipage} \\
\midrule\noalign{}
\endhead
\bottomrule\noalign{}
\endlastfoot
\(\varepsilon_w\) & Within-segment variety-level labor-demand
elasticities, or desired unsubsidized wage markups under the maintained
assumptions. \\
\(\nu\) & Relative labor-input responses to relative wages across the
modeled wage-setting segments. \\
\(\delta\) & Difference in the segments' hand-to-mouth shares,
\(q_1-q_2\), measured from joint liquidity-status and segment data. \\
\(\eta_w\) & Wage-inflation responses to wage-setting incentives and the
dynamics of segment wages. \\
\(\lambda\) & Household share consistent with current-income consumption
and no asset-market participation. \\
\end{longtable}

\emph{Notes:} These are potential moments, not estimates used in this
paper. Segment definitions and household classifications must match the
maintained membership and population restrictions. Occupation, industry,
or contract categories are candidate segment definitions rather than
direct measures of financial type.

Relative wages and labor input are jointly endogenous, so identifying
\(\nu\) requires distinguishing movements along labor-demand schedules
from shifts in segment demand. The alignment parameter measures a
difference in conditional household shares, not an unrestricted
correlation between liquidity status and occupation. Wage-adjustment
moments inform \(\eta_w\) jointly with \(\varepsilon_w\), preferences,
and shock dynamics: the wage-equation slope is
\(\kappa_w=\varepsilon_w/\eta_w\), while observed persistence also
reflects equilibrium forcing. The first-order wage equations contain
\(\varepsilon_w\) and \(\eta_w\) through this ratio, so their slopes
alone do not identify the two parameters separately. The quadratic
Rotemberg cost has no direct mapping to an observed wage-reset
frequency. The present calibration does not match these moments; the
sensitivity grid therefore describes model dependence rather than an
empirically validated parameter region.

The implied reduced-form coefficients summarize the adjustment dynamics
in the aggregate and distributional wage blocks. With the benchmark
calibration, \(\Theta_\sigma=5.99\) in the wage-gap equation and
\(\Theta_w=2.39\) in the aggregate real-wage equation.
Table~\ref{tbl-wage-roots} reports the characteristic roots. The
wage-gap equation has a stable root of \(0.172\), while its unstable
root governs the discounted forward solution.

\begin{table}[H]

\caption{\label{tbl-wage-roots}Wage-block roots}

\centering{

\centering
\begin{tabular}[t]{>{\raggedright\arraybackslash}p{5.0cm}ccc}
\toprule
Block & Composite coefficient & Stable root & Unstable root\\
\midrule
Cross-type wage gap, $\sigma_t^w$ & $\Theta_\sigma=5.99$ & $0.172$ & $5.879$\\
Aggregate real wage, $w_t$ & $\Theta_w=2.39$ & $0.539$ & $1.876$\\
\bottomrule
\end{tabular}

}

\end{table}%

Output gaps, price inflation, and cross-type wage and consumption gaps
are reported as log deviations. The profit share, transfers, and their
difference---the net distributional wedge---are normalized by
steady-state output. In the monetary experiment, \(e_{m,0}=0.01\) is the
impact innovation at horizon \(h=0\). All subsequent innovations are
zero, and the monetary state decays at its calibrated AR(1) rate. The
IRFs report 20 quarters, and cumulative statistics sum
\(h=0,\ldots,19\). The innovation \(e_{m,0}=0.01\) shifts the quarterly
log gross policy-rate rule down by \(0.01\): 100 basis points in
log-rate units, or 400 basis points when annualized. This is the
exogenous rule shift, not the realized interest-rate response, which
also includes endogenous inflation feedback.

The main text compares three economic benchmarks. Type-Specific-Wage
TANK is the full heterogeneous-agent model with type-specific sticky
wages and sticky prices. The reduced-form common-wage closure retains
the aggregate sticky-wage equation. It removes the Type H wage-setting
block and the reallocation of employment through relative wages; CES
labor demand then implies equal employment across types. Dual-Rigidity
RANK sets \(\lambda=0\), thereby removing household heterogeneity.
Appendix C.2 reports Sticky-Price-Only TANK. This specification sets
wage-adjustment costs to zero while keeping \(\psi_w\) finite, making
the wage gap static rather than inertial.
Table~\ref{tbl-wage-specifications} covers the two specifications used
in the cross-closure comparison; RANK and Sticky-Price-Only TANK are
auxiliary parameter-limit benchmarks rather than additional wage-setting
closures.

I solve the canonical system in MATLAB/Dynare. The dynamic canonical
coefficients above are defined for \(\eta_w>0\). For Sticky-Price-Only
TANK, I use the static flexible-wage relations for the \(\eta_w=0\)
limit, as derived in Section 2. I impose Dual-Rigidity RANK by setting
\(\lambda=0\). Under this restriction, \(\sigma_t^w\) and \(\sigma_t^c\)
are identically zero, so their distributional moments are omitted where
appropriate.

Appendix C.1 reports local-determinacy diagnostics for the
passive-transfer system. The reduced-form common-wage comparison is the
central cross-closure experiment below; the online appendix provides
supplementary diagnostics for inflation and consumption-gap persistence.
Appendix C.3 presents baseline-centered sensitivity checks for
\(\eta_w\), \(\lambda\), and \(\psi_w\). The replication package
contains the MATLAB/Dynare scripts and generated CSV series underlying
the figures, tables, and joint-grid diagnostics, together with an
exact-arithmetic verification of the benchmark impact result.

The additional experiments use a stable Schur-subspace solution checked
against the baseline Dynare solution, with equilibrium residuals and
limiting cases reported in the replication materials.

\subsection{Equilibrium Impact
Attenuation}\label{equilibrium-impact-attenuation}

\textbf{Benchmark result.} For the calibration in
Table~\ref{tbl-calibration} and under the passive-transfer rule, a
one-time monetary, transfer, or technology innovation from the steady
state satisfies the condition in
Proposition~\ref{prp-impact-attenuation}, stated in
(\ref{eq-equilibrium-offset-condition}). Thus, for each shock
individually, earnings reallocation provides a strict partial offset to
direct incidence on impact. Table~\ref{tbl-equilibrium-impact} reports
the multipliers and offset shares from the full-equilibrium calculation.

\begin{table}[H]

\caption{\label{tbl-equilibrium-impact}Equilibrium impact offsets}

\centering{

\centering
\begin{tabular}[t]{lrr}
\toprule
Shock & Discounted-wedge multiplier & Impact offset (\%)\\
\midrule
Monetary & 1.257111 & 38.88\\
Transfer & 1.159099 & 35.85\\
Technology & 1.108821 & 34.29\\
\bottomrule
\end{tabular}

}

\end{table}%

\emph{Notes:} The multiplier is \(\mathcal M_{0,i}^\omega\) and the
offset share is \(100\mathcal B_{0,i}^{\mathrm{GE}}\) for an innovation
of type \(i\) when the inherited states are zero. The calculation uses
(\ref{eq-equilibrium-resolvent}) to sum the full equilibrium's expected
wedge over an infinite horizon. The shares are invariant to the
magnitude and sign of an individual nonzero innovation. They compare
direct and net incidence within the same equilibrium.

This result evaluates Proposition~\ref{prp-impact-attenuation} in the
full equilibrium; it does not impose the AR(1)-type wedge restriction in
Corollary~\ref{cor-ar1-wedge}. The benchmark disturbances have a common
persistence of \(\rho_\star=4/5\), and the system has two additional
stable endogenous roots, \(\zeta_+=0.46417381\ldots\) and
\(\zeta_-=0.17393030\ldots\). For each type of innovation, the wedge
response takes the form
\begin{equation}\protect\phantomsection\label{eq-equilibrium-wedge-modes}{
\omega_{h,i}=\varepsilon_{i,0}
\left(
\vartheta_{\star i}^\omega\rho_\star^h
+\vartheta_{+i}^\omega\zeta_+^h
+\vartheta_{-i}^\omega\zeta_-^h
\right),\qquad h\geq0,
}\end{equation} where all three loadings are nonzero and normalized to a
unit innovation. Summing each geometric component exactly gives
\begin{equation}\protect\phantomsection\label{eq-equilibrium-modal-multiplier}{
\mathcal M_{0,i}^\omega
=
\frac{
\dfrac{\vartheta_{\star i}^\omega}{1-\rho_\star/\xi_2}
+\dfrac{\vartheta_{+i}^\omega}{1-\zeta_+/\xi_2}
+\dfrac{\vartheta_{-i}^\omega}{1-\zeta_-/\xi_2}
}{
\vartheta_{\star i}^\omega+\vartheta_{+i}^\omega+\vartheta_{-i}^\omega
}.
}\end{equation} Appendix B.3 derives the algebraic roots and loadings
and certifies rational bounds strictly inside \((0,1)\) for all three
offset shares. This provides a computer-assisted exact verification of
the condition in Proposition~\ref{prp-impact-attenuation}, rather than
merely a check based on rounded numerical responses. The proof applies
specifically to the benchmark calibration and individual structural
innovations; it is not a uniform result over the parameter space.
Later-horizon accounting shares also reflect the inherited term in
(\ref{eq-equilibrium-inherited-decomposition}).

The extended solver independently reproduces these benchmark offsets
using the full stable transition and agrees with the Dynare responses.
Its numerical checks are distinct from the rational-arithmetic
certificate above.

\subsection{Benchmark Monetary Experiment: Accounting Offset and Closure
Contrast}\label{benchmark-monetary-experiment-accounting-offset-and-closure-contrast}

The first exercise holds the type-specific-wage equilibrium fixed and
decomposes its consumption gap. The absolute gap reaches its maximum at
\(h=1\). At this horizon, with all terms measured in log deviations,
(\ref{eq-static-dispersion}) gives \[
\underbrace{-0.00872}_{\text{direct profit-transfer incidence}}
+
\underbrace{0.00368}_{\text{earnings-reallocation buffer}}
=
\underbrace{-0.00504}_{\text{net consumption gap}}.
\] In the nested perfect-alignment benchmark, the earnings-reallocation
term therefore offsets 42.2 percent of the direct component at that
horizon. Over \(h=0,\ldots,19\), the corresponding within-path
accounting offset is 43.9 percent. These shares do not measure
reductions relative to the common-wage economy. Across the 36-cell joint
grid described below, the peak accounting offset ranges from 10.9 to
47.2 percent. The cumulative offset ranges from 17.6 to 47.3 percent.
Both remain strictly between zero and 100 percent.

\begin{figure}[H]

\includegraphics[width=1\linewidth,height=\textheight,keepaspectratio]{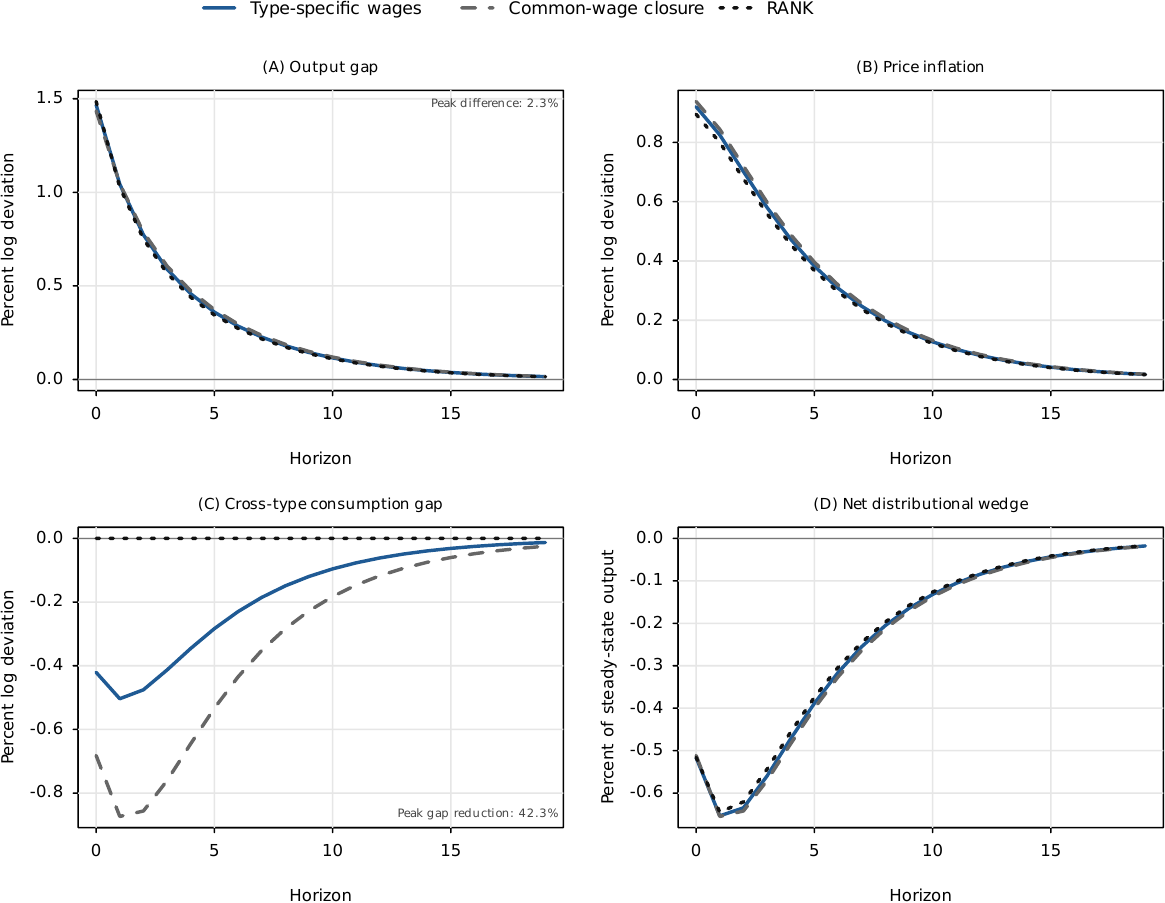}

\par\smallskip
\begin{minipage}{0.98\linewidth}
\raggedright
\footnotesize \textit{Notes:} Deterministic first-order responses at the nested perfect-alignment benchmark to $e_{m,0}=0.01$ over horizons $h=0,\ldots,19$, with all subsequent innovations set to zero. Panels report the output gap, price inflation, the cross-type consumption gap, and the net distributional wedge for Type-Specific-Wage TANK, the Reduced-Form Common-Wage Closure, and Dual-Rigidity RANK. Panels (A)--(C) use percent log deviations; panel (D) uses percent of steady-state output.
\end{minipage}

\caption{\label{fig-common-wage-monetary}Monetary responses}

\end{figure}

The second exercise compares two general-equilibrium closures.
Figure~\ref{fig-common-wage-monetary} compares responses from the
type-specific-wage model, the reduced-form common-wage closure, and the
RANK benchmark. At the benchmark, the peak output gap is 0.01468 (about
1.47 percent) with type-specific wages and 0.01435 (about 1.44 percent)
under the common-wage closure, a difference of 2.3 percent. The
cumulative output responses over \(h=0,\ldots,19\) are 0.06044 and
0.06126 log-deviation quarters, respectively, making the
type-specific-wage response 1.3 percent smaller.

The difference in consumption incidence across the two closures is
larger. At the nested perfect-alignment benchmark, the maximum absolute
consumption gap is 0.00504 (0.504 log percent) with type-specific wages
and 0.00873 (0.873 log percent) under the common-wage closure, a
difference of 42.3 percent. The corresponding cumulative absolute gaps
are 0.03555 and 0.06467 log-deviation quarters, a difference of 45.0
percent. These figures compare the baseline equilibrium with the
reduced-form common-wage closure and include the resulting
general-equilibrium adjustment; they do not represent the causal effect
of a single structural parameter. The numerical proximity of the 42.2
percent accounting offset and the 42.3 percent cross-closure difference
is a coincidence at the benchmark, not an identity.

The direction of the cross-closure difference in the consumption gap
persists throughout the joint sensitivity exercise. The grid combines
\(\lambda\in\{0.10,0.25,0.40\}\), \(\psi_w\in\{2,5,10,20\}\), and
\(\eta_w\in\{25,50,100\}\). Both the baseline own-lag specification and
the reduced-form common-wage closure are determinate in all 36 cells.
Across the grid, the peak consumption gap with type-specific own-lag
wages is 10.9--47.8 percent smaller than that under the common-wage
closure. The incidence ordering holds in all 36 cells. The peak output
difference ranges from 0.06 to 9.0 percent and remains within 5 percent
in 30 cells, so aggregate transmission is often similar but not
uniformly so across the grid.

Three cells hold \(\lambda=0.25\) and \(\kappa_w=0.2\) fixed while
varying \((\psi_w,\eta_w)\) among \((5,25)\), \((10,50)\), and
\((20,100)\). The peak accounting offset is 35.3, 42.2, and 46.0
percent, respectively; the corresponding cross-closure reductions in the
peak consumption gap are 35.2, 42.3, and 46.2 percent. Fixing the slope
of the wage equation therefore does not fix distributional incidence:
labor substitutability also enters relative earnings and the wage-gap
adjustment coefficient. This comparison jointly varies the two primitive
parameters and illustrates model sensitivity without identifying an
empirically admissible range.

\subsection{Inherited State and Expected-Wedge
Forcing}\label{inherited-state-and-expected-wedge-forcing}

The third exercise decomposes the stable solution for the wage gap. At
the peak of the benchmark monetary response, the inherited component
accounts for 17.0 percent of \(|\sigma_t^w|\); over the full response
horizon, it accounts for 17.1 percent of the two components' combined
absolute contributions. The conditional loading on the previous wage gap
is \(\xi_1=0.172\), implying a half-life of 0.39 quarters when the
expected wedge path is held fixed. For a general-equilibrium transition
driven solely by the inherited state, the half-peak threshold is crossed
after one quarter.

The full wage-gap response to the monetary shock is more persistent
because it combines this state with equilibrium forcing. The wage gap
reaches its trough at horizon 2 and has a post-peak half-life of five
quarters, as does the equilibrium wedge. Expected-wedge forcing
therefore accounts for most of the benchmark path. The five-quarter
impulse-response half-life is not, by itself, a measure of own-lag
persistence.

This benchmark conclusion does not hold uniformly over the joint grid.
The cumulative inherited-state share ranges from 2.9 to 74.9 percent,
the conditional half-life from 0.20 to 2.50 quarters, and the full
monetary wage-gap half-life from four to eight quarters. The inherited
state is secondary at the benchmark, not throughout the reported
parameter space.

A supplementary reduced-form timing diagnostic replaces the
type-specific lagged wage references with a common lagged wage index
while retaining the original forward-looking wage Phillips curves.
Appendix A.2 defines this diagnostic; it differs from the companion
paper's aggregate-lag optimization problem and static relative-wage
mapping. Relative to this diagnostic, own-lag adjustment raises the peak
absolute consumption gap by 4.5 percent and changes the sum over
\(h=0,\ldots,19\) by \(-0.3\) percent. The path distance equals 4.9
percent of the diagnostic's cumulative gap. This comparison changes the
forcing map and the equilibrium wedge path as well as removing the
inherited state, so it is not a pure component decomposition or a
comparison of two optimized adjustment-cost models. The exact
decomposition assigns a nonzero role to the state, while the diagnostic
mainly changes the timing of benchmark incidence.

\begin{figure}[H]

\centering{

\includegraphics[width=1\linewidth,height=\textheight,keepaspectratio]{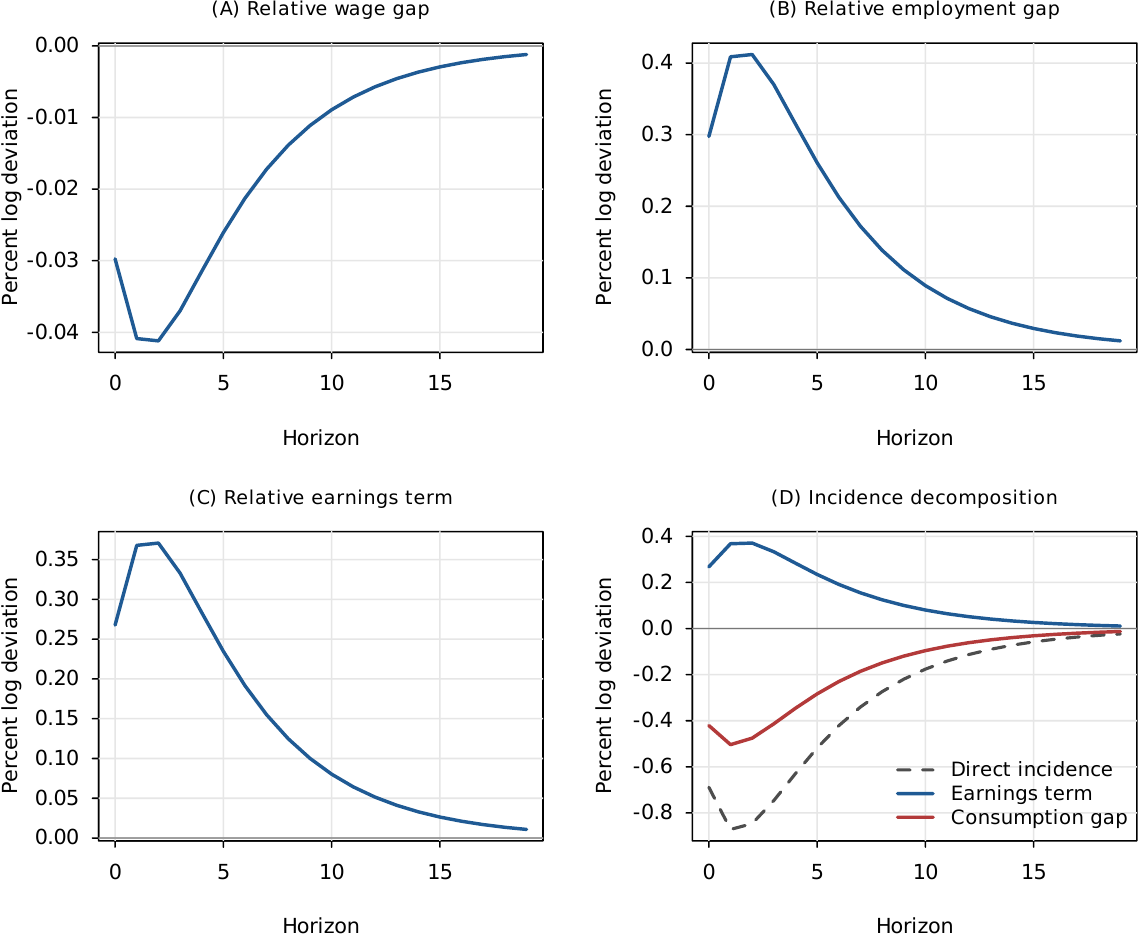}

\par\smallskip
\begin{minipage}{0.98\linewidth}
\raggedright
\footnotesize \textit{Notes:} Deterministic first-order responses in Type-Specific-Wage TANK to $e_{m,0}=0.01$ over horizons $h=0,\ldots,19$, with all subsequent innovations set to zero. Relative gaps are Type S minus Type H. All panels show percent log deviations, including the direct-incidence and earnings contributions to the consumption gap. The earnings term is $(1-\psi_w)\sigma_t^w$; direct incidence is $\omega_t/(1-\lambda)$; and their sum is $\sigma_t^c$.
\end{minipage}

}

\caption{\label{fig-monetary-shock}Distributional-buffer mechanism}

\end{figure}

Figure~\ref{fig-monetary-shock} traces the full labor-market path rather
than the inherited component alone. The monetary expansion lowers the
Type-S relative wage and shifts labor demand toward Type S. With
\(\psi_w=10\), the relative employment response is ten times the wage
response in absolute value, so the relative-earnings term is positive.
Relative employment and earnings follow \(\sigma_t^n=-\psi_w\sigma_t^w\)
and \(\sigma_t^{w+n}=(1-\psi_w)\sigma_t^w\). Panel (D) shows how the
resulting earnings term offsets part of the negative direct-incidence
term.

\subsection{Scope Across Shocks and
Parameters}\label{scope-across-shocks-and-parameters}

The within-path accounting offset also arises under transfer and
technology shocks. Across the monetary, transfer, and technology
experiments, earnings reallocation offsets 34--42 percent of direct
incidence at the peak and 42--44 percent on the 20-quarter cumulative
absolute measure. The technology shock reverses the signs of both terms:
direct incidence is positive, whereas earnings reallocation is negative.
The accounting offset is therefore not tied to a particular sign of
direct incidence. Appendix C.2 and
Table~\ref{tbl-cross-shock-accounting} provide the underlying
calculations. These are accounting statements along realized
type-specific-wage paths, not cross-closure reductions or welfare
rankings.

The one-parameter wage-rigidity sweep under the monetary shock changes
both aggregate and distributional dynamics. As shown in the monetary
wage-adjustment-cost sweep in the online appendix, raising \(\eta_w\)
from \(10\) to \(150\) more than doubles the peak output-gap response
but lowers the maximum consumption-gap response. In the
type-specific-wage model, this sweep has the same directional ordering
as the common-wage result in Bojovic and Ellis (2026) and the opposite
ordering from the labor-income-only benchmark in Diz et al. (2023).
Here, greater wage flexibility is associated with a smaller peak
absolute output-gap response. The comparison is only directional because
the present sweep varies a Rotemberg wage-adjustment cost and reports
the peak response to a monetary-policy shock. The cited studies instead
vary Calvo common-wage rigidity and examine output volatility under
demand shocks.

The 42.3 percent cross-closure difference in the nested
perfect-alignment benchmark is conditional on \(\delta=1\),
\(\nu=\varepsilon_w=10\), and the maintained disposable-income block. In
this block, hand-to-mouth households receive transfers but no profits,
carry no debt, and have the same productivity as savers. The richer
income and balance-sheet specification in Bojovic and Ellis (2026) can
change the aggregate effects of common-wage flexibility, but it neither
introduces nor evaluates the relative-earnings buffer studied here.

The joint grid and wage-rigidity sweep show that the incidence ordering
across closures is stable over the reported grid, but that the
similarity in aggregate transmission is not uniform. The accounting
decomposition measures the buffer along a given equilibrium path; the
common-wage comparison evaluates a different equilibrium closure; and
the stable-solution decomposition separates inherited dynamics from
expected-wedge forcing.

\subsection{Separate Elasticities and Financial
Alignment}\label{separate-elasticities-and-financial-alignment}

The extension holds the individual wage-setting elasticity at
\(\varepsilon_w=10\) and the wage-adjustment cost at \(\eta_w=50\),
while varying \(\nu\in\{0.5,1,2,5,10,20\}\) and
\(\delta\in\{0,0.25,0.5,0.75,1\}\). The 30 parameter combinations are
determinate, and each is evaluated for monetary, transfer, and
technology shocks. Unlike the original \(\psi_w\) sweep, changing
\(\nu\) here leaves the wage Phillips-curve slope and the desired
individual wage markup fixed. The table includes partial alignment at
\(\delta=0.5\) with \(\nu=2\) and \(5\), alongside the nested
perfect-alignment benchmark; these alternatives are sensitivity cases,
not empirical calibrations.

\Needspace{.58\textheight}

{

\begin{longtable}[]{@{}
  >{\raggedleft\arraybackslash}p{(\linewidth - 8\tabcolsep) * \real{0.2000}}
  >{\raggedleft\arraybackslash}p{(\linewidth - 8\tabcolsep) * \real{0.1800}}
  >{\raggedleft\arraybackslash}p{(\linewidth - 8\tabcolsep) * \real{0.2000}}
  >{\raggedleft\arraybackslash}p{(\linewidth - 8\tabcolsep) * \real{0.1900}}
  >{\raggedleft\arraybackslash}p{(\linewidth - 8\tabcolsep) * \real{0.2300}}@{}}

\caption{\label{tbl-segmentation-incidence}Substitution and alignment}

\tabularnewline

\toprule\noalign{}
\begin{minipage}[b]{\linewidth}\raggedleft
Substitution
\end{minipage} & \begin{minipage}[b]{\linewidth}\raggedleft
Alignment
\end{minipage} & \begin{minipage}[b]{\linewidth}\raggedleft
Impact offset (\%)
\end{minipage} & \begin{minipage}[b]{\linewidth}\raggedleft
Peak offset (\%)
\end{minipage} & \begin{minipage}[b]{\linewidth}\raggedleft
Peak gap reduction (\%)
\end{minipage} \\
\midrule\noalign{}
\endhead
\bottomrule\noalign{}
\endlastfoot
0.50 & 1.00 & -19.17 & -30.70 & -26.80 \\
1.00 & 1.00 & 0.00 & 0.00 & 0.00 \\
2.00 & 0.50 & 4.57 & 5.59 & 5.44 \\
2.00 & 1.00 & 15.76 & 19.02 & 18.70 \\
5.00 & 0.50 & 10.39 & 12.04 & 11.94 \\
5.00 & 1.00 & 31.15 & 35.30 & 35.25 \\
10.00 & 0.00 & 0.00 & 0.00 & 0.00 \\
10.00 & 0.25 & 3.90 & 4.37 & 4.35 \\
10.00 & 0.50 & 13.90 & 15.44 & 15.43 \\
10.00 & 0.75 & 26.50 & 29.11 & 29.16 \\
10.00 & 1.00 & 38.88 & 42.21 & 42.33 \\
20.00 & 1.00 & 43.81 & 46.02 & 46.24 \\

\end{longtable}

}

\emph{Notes:} Monetary innovation \(e_{m,0}=0.01\) from zero inherited
states. Substitution is \(\nu\) and alignment is \(\delta\); the nested
perfect-alignment benchmark is \((\nu,\delta)=(10,1)\). The impact and
peak offsets compare earnings reallocation with direct incidence within
the segment-wage equilibrium; the peak offset is evaluated at the date
of its maximum absolute consumption gap. The last column compares
maximum absolute gaps across the segment-wage and common-wage economies
over \(h=0,\ldots,19\), using the common-wage maximum as denominator. A
negative value denotes amplification or a larger gap, respectively. All
rows hold the other benchmark parameters fixed.

With perfect alignment and \(\nu=0.5\), earnings reallocation amplifies
monetary impact incidence by 19.17 percent, and the peak consumption gap
is 26.80 percent larger than under the common-wage closure. At
\(\nu=1\), the earnings term vanishes. For \(\nu=10\), the impact offset
rises from zero at identical segment composition to 3.90, 13.90, 26.50,
and 38.88 percent as alignment increases. Thus both the sign reversal
and the zero-alignment endpoint appear in the full-equilibrium
solutions. Monotonicity over these solved cells does not establish a
general-equilibrium comparative-static theorem outside the reported
grid.

At \(\delta=0.5\), the peak consumption-gap reduction is 5.44 percent
for \(\nu=2\) and 11.94 percent for \(\nu=5\), compared with 42.33
percent at \((\nu,\delta)=(10,1)\). Their impact accounting offsets are
4.57 and 10.39 percent, respectively. The roughly two-fifths benchmark
reduction therefore depends materially on the assumed alignment and
substitution. A markup calibration for individual wage setters does not
identify either of those two parameters.

\subsection{Aggregate-IRF Matching}\label{aggregate-irf-matching}

The original comparison holds all coefficients fixed. A second exercise
chooses the common-wage adjustment cost to fit the nested
perfect-alignment benchmark's output-gap and price-inflation responses
to the monetary shock. The target retains \(\delta=1\) and
\(\nu=\varepsilon_w=10\). It minimizes
\begin{equation}\protect\phantomsection\label{eq-irf-matching-loss}{
L(\eta_w^{CW})=
\sum_{j\in\{x,\pi^p\}}
\frac{\sum_{h=0}^{19}[j_h^{CW}(\eta_w^{CW})-j_h^{TS}]^2}
{\sum_{h=0}^{19}(j_h^{TS})^2},
\qquad \eta_w^{CW}\in[1,500].
}\end{equation} The price-adjustment cost remains 50, other parameters
and the innovation remain fixed, and admissible fits must be
determinate. A logarithmic grid, local bracket refinements, the
endpoints, and the original calibration are compared. This is a
constrained one-parameter approximation exercise; it does not fit
consumption responses or estimate parameters from data.

\Needspace{.34\textheight}

{

\begin{longtable}[]{@{}
  >{\raggedright\arraybackslash}p{(\linewidth - 8\tabcolsep) * \real{0.1667}}
  >{\raggedleft\arraybackslash}p{(\linewidth - 8\tabcolsep) * \real{0.1282}}
  >{\raggedleft\arraybackslash}p{(\linewidth - 8\tabcolsep) * \real{0.2179}}
  >{\raggedleft\arraybackslash}p{(\linewidth - 8\tabcolsep) * \real{0.2564}}
  >{\raggedleft\arraybackslash}p{(\linewidth - 8\tabcolsep) * \real{0.2308}}@{}}

\caption{\label{tbl-irf-matching}Aggregate fit and consumption
incidence}

\tabularnewline

\toprule\noalign{}
\begin{minipage}[b]{\linewidth}\raggedright
Comparison
\end{minipage} & \begin{minipage}[b]{\linewidth}\raggedleft
Wage cost
\end{minipage} & \begin{minipage}[b]{\linewidth}\raggedleft
Output error (\%)
\end{minipage} & \begin{minipage}[b]{\linewidth}\raggedleft
Inflation error (\%)
\end{minipage} & \begin{minipage}[b]{\linewidth}\raggedleft
Gap reduction (\%)
\end{minipage} \\
\midrule\noalign{}
\endhead
\bottomrule\noalign{}
\endlastfoot
Original & 50.00 & 2.21 & 2.67 & 42.33 \\
Monetary fit & 51.31 & 2.39 & 2.26 & 41.88 \\
Joint fit & 47.40 & 3.01 & 3.53 & 43.24 \\

\end{longtable}

}

\emph{Notes:} Every row evaluates the monetary responses. Errors are 100
times the square root of each normalized squared-error term in
(\ref{eq-irf-matching-loss}); they are relative path-norm errors, not
peak differences. Gap reduction compares maximum absolute consumption
gaps using the fitted common-wage gap as denominator. The joint fit
averages the same loss over monetary, transfer, and technology
responses. All other primitives and shock normalizations are held fixed.

The monetary fit selects \(\eta_w^{CW}=51.31\) and reduces the combined
normalized squared loss from 0.0012005 to 0.0010825. Its output and
inflation path errors are 2.39 and 2.26 percent, while the peak
consumption gap remains 41.88 percent smaller in the nested
perfect-alignment benchmark. Aggregate fitting therefore leaves a
substantial incidence difference under that calibration; it does not
establish that this magnitude survives weaker alignment or lower segment
substitution. The fit trades off the two aggregate targets: inflation
error falls, whereas output error rises slightly. It does not improve
each target separately.

The joint fit selects \(\eta_w^{CW}=47.40\). Its monetary comparison
also retains a large consumption-gap difference, but the transfer
responses are not closely matched: their output and inflation path
errors are about 48 and 82 percent. Thus the approximation result
remains shock specific. The online appendix reports all fitted-shock
evaluations.

\subsection{Persistence, Policy, and Price
Adjustment}\label{persistence-policy-and-price-adjustment}

A further set of one-parameter-at-a-time experiments varies each shock's
own persistence over \(\{0,0.5,0.8,0.95\}\), the Taylor coefficient over
\(\{1.1,1.5,2\}\), and the price-adjustment cost over \(\{25,50,100\}\).
The latter changes \(\kappa_p\) while holding \(\psi_p=10\). Each
variation is evaluated for all three shocks and both closures. All 30
reported experiment rows are determinate. The ranges are sensitivity
checks around the illustrative benchmark, not empirically admissible
parameter intervals.

\Needspace{.46\textheight}

{

\begin{longtable}[]{@{}
  >{\raggedright\arraybackslash}p{(\linewidth - 6\tabcolsep) * \real{0.1594}}
  >{\raggedright\arraybackslash}p{(\linewidth - 6\tabcolsep) * \real{0.2464}}
  >{\raggedleft\arraybackslash}p{(\linewidth - 6\tabcolsep) * \real{0.2609}}
  >{\raggedleft\arraybackslash}p{(\linewidth - 6\tabcolsep) * \real{0.3333}}@{}}

\caption{\label{tbl-additional-sensitivity}Shock and policy sensitivity}

\tabularnewline

\toprule\noalign{}
\begin{minipage}[b]{\linewidth}\raggedright
Shock
\end{minipage} & \begin{minipage}[b]{\linewidth}\raggedright
Variation
\end{minipage} & \begin{minipage}[b]{\linewidth}\raggedleft
Impact offset (\%)
\end{minipage} & \begin{minipage}[b]{\linewidth}\raggedleft
Peak gap reduction (\%)
\end{minipage} \\
\midrule\noalign{}
\endhead
\bottomrule\noalign{}
\endlastfoot
Monetary & Persistence & 33.59 to 40.06 & 35.04 to 44.27 \\
Monetary & Taylor response & 38.69 to 39.10 & 42.02 to 43.57 \\
Monetary & Price adjustment & 38.30 to 39.39 & 42.29 to 44.12 \\
Transfer & Persistence & 31.02 to 36.89 & 32.42 to 36.67 \\
Transfer & Taylor response & 35.83 to 35.90 & 36.11 to 37.15 \\
Transfer & Price adjustment & 35.82 to 35.89 & 36.12 to 36.84 \\
Technology & Persistence & 30.87 to 34.29 & 32.27 to 35.46 \\
Technology & Taylor response & 33.95 to 34.84 & 34.96 to 36.36 \\
Technology & Price adjustment & 33.64 to 34.93 & 34.57 to 36.35 \\

\end{longtable}

}

\emph{Notes:} Impact offset is within-equilibrium accounting; peak gap
reduction is the cross-closure comparison. Each shock has a 0.01
innovation in its own units. Persistence changes only for the named
shock, with all other parameters at baseline. Peak responses use 20
quarters. The table reports extrema over each specified sweep, including
repeated benchmark cells.

The impact accounting offsets range from 30.87 to 40.06 percent, and
cross-closure peak-gap reductions range from 32.27 to 44.27 percent.
Persistence changes the discounted wedge directly and also changes its
endogenous propagation. Taylor feedback and price rigidity alter that
propagation even though neither changes the consumption-gap identity.
These results extend the original determinacy diagnostics to the size of
distributional incidence.

\section{Conclusion}\label{conclusion}

Relative wages transmit profit-transfer incidence through employment and
earnings. The consumption-gap identity separates this channel from
direct incidence, and the stable wage-gap solution characterizes when
the discounted expected wedge produces a partial offset. Own-lag
adjustment adds an inherited state, whose role is distinct from the
persistence of equilibrium forcing.

The sign of reallocation depends on labor-market structure. Separating
the individual wage-setting elasticity from substitution between
segments admits both attenuation and amplification. In the mixed-segment
extension, identical financial-type composition removes the
relative-earnings channel; perfect alignment recovers the benchmark
first-order system. These statements depend on the specified
wage-setting and income-pooling arrangements rather than identifying a
universal effect of liquidity status.

The nested perfect-alignment monetary benchmark has a peak consumption
gap about two-fifths smaller than the common-wage closure despite
similar peak output responses. This magnitude depends on the imposed
alignment and segment substitution: with \(\delta=0.5\), the reduction
is 5.44 percent at \(\nu=2\) and 11.94 percent at \(\nu=5\), holding the
individual wage-setting elasticity at ten. The additional experiments
assess aggregate-IRF matching, shock and policy sensitivity, and partial
alignment. Transfer comparisons distinguish a fixed per-recipient
payment from a fixed aggregate payment because their comparative statics
in the hand-to-mouth share differ. None of these consumption-gap
comparisons is a welfare ranking.

An empirical application requires joint information on liquidity status,
wage-setting segments, relative wages, employment, and earnings. Those
moments would discipline the substitution and alignment parameters
separately. The analytical contribution is to identify the conditions
under which relative earnings buffer unequal exposure and the conditions
under which that conclusion changes.

\section*{References}\label{references}
\addcontentsline{toc}{section}{References}

\protect\phantomsection\label{refs}

\section*{Appendix}\label{appendix}
\addcontentsline{toc}{section}{Appendix}

\subsection*{Appendix A. Algebraic Derivations and the Common-Wage
Closure}\label{appendix-a.-algebraic-derivations-and-the-common-wage-closure}
\addcontentsline{toc}{subsection}{Appendix A. Algebraic Derivations and
the Common-Wage Closure}

\subsubsection*{A.1 Algebraic Derivation of the Canonical
Representation}\label{a.1-algebraic-derivation-of-the-canonical-representation}
\addcontentsline{toc}{subsubsection}{A.1 Algebraic Derivation of the
Canonical Representation}

The main text presents the nonredundant 20-equation system and the
canonical representation in Section 2. This appendix provides the
underlying analytical reduction.

\textbf{1. Static Consumption-Gap Block.} The derivation first expresses
Type H consumption relative to the aggregate mean. Subtracting the
aggregate resource identity \(c_t = w_t + n_t + d_t\) from
\(c^H_t = w^H_t + n^H_t + z_t\) and using \(d_t = a_t - w_t\) yields
\[ \tilde{c}^H_t = \tilde{w}^H_t + \tilde{n}^H_t + z_t - (a_t - w_t) \]
Type H labor demand in deviation form is
\(\tilde{n}^H_t = -\psi_w \tilde{w}^H_t\), so
\[ \tilde{c}^H_t = (1-\psi_w)\tilde{w}^H_t + z_t - (a_t - w_t) \] The
cross-sectional mappings \(\tilde{w}^H_t = -(1-\lambda)\sigma^w_t\) and
\(\tilde{c}^H_t = -(1-\lambda)\sigma^c_t\), together with the net
distributional wedge \(\omega_t = a_t - w_t - z_t\), then give
\[ -(1-\lambda)\sigma^c_t = -(1-\psi_w)(1-\lambda)\sigma^w_t - \omega_t \implies \sigma^c_t = (1-\psi_w)\sigma^w_t + \frac{\omega_t}{1-\lambda} \]

\textbf{2. Wage-Gap Dynamics.} The next step isolates the type-specific
wage component. Subtracting the aggregate wage Phillips curve from the
Type H wage Phillips curve gives:
\[ \eta_w (\pi^H_t - \pi^w_t) = \beta \eta_w \mathbb{E}_t (\pi^H_{t+1} - \pi^w_{t+1}) - \psi_w (\mu^H_t - \mu^w_t) \]
The identities \(\pi^H_t - \pi^w_t = \tilde{w}^H_t - \tilde{w}^H_{t-1}\)
and \(\tilde{w}^H_t = -(1-\lambda)\sigma^w_t\) express wage inflation in
terms of the wage gap. The markup gap is
\[ \mu^H_t - \mu^w_t = \tilde{w}^H_t - \gamma \tilde{c}^H_t - \varphi \tilde{n}^H_t \]
Using \(\tilde{n}^H_t = -\psi_w \tilde{w}^H_t\) and
\(\tilde{c}^H_t = (1-\psi_w)\tilde{w}^H_t - \omega_t\) gives \[
\begin{aligned}
\mu^H_t-\mu^w_t
&=\tilde w_t^H-\gamma[(1-\psi_w)\tilde w_t^H-\omega_t]+\varphi\psi_w\tilde w_t^H\\
&=[1-\gamma(1-\psi_w)+\varphi\psi_w]\tilde w_t^H+\gamma\omega_t.
\end{aligned}
\] Combining this markup gap with
\(\tilde{w}^H_t = -(1-\lambda)\sigma^w_t\) and the wage-Phillips-curve
difference yields: \[
\begin{aligned}
-\eta_w(1-\lambda)(\sigma_t^w-\sigma_{t-1}^w)
={}&-\beta\eta_w(1-\lambda)\mathbb{E}_t(\sigma_{t+1}^w-\sigma_t^w)\\
&-\psi_w\left[-(1-\lambda)[1-\gamma(1-\psi_w)+\varphi\psi_w]\sigma_t^w+\gamma\omega_t\right].
\end{aligned}
\] Dividing by \(-\eta_w(1-\lambda)\), using
\(\kappa_w \equiv \psi_w/\eta_w\), and rearranging terms yields \[
-\beta \mathbb{E}_t \sigma^w_{t+1}
+
\Theta_\sigma \sigma_t^w
-
\sigma_{t-1}^w
=
\frac{\gamma\kappa_w}{1-\lambda}\omega_t,
\] where \[
\Theta_\sigma \equiv 1+\beta+\kappa_w[(1-\gamma)+\psi_w(\gamma+\varphi)].
\] This is the wage-gap law used in
Section~\ref{sec-perfect-alignment-main}.

\textbf{3. The Macroeconomic Supply Block.} This block derives the
aggregate real-wage equation that closes the supply side of the
canonical system. The key step combines the price and wage Phillips
curves with the real-wage inflation identity. Define
expectations-adjusted price and wage inflation by
\(\hat{\pi}^p_t = \pi^p_t - \beta \mathbb{E}_t \pi^p_{t+1}\) and
\(\hat{\pi}^w_t = \pi^w_t - \beta \mathbb{E}_t \pi^w_{t+1}\),
respectively. The aggregate Phillips curves imply:
\[ \hat{\pi}^p_t = -\frac{\psi_p}{\eta_p}(a_t - w_t) \quad \text{and} \quad \hat{\pi}^w_t = -\frac{\psi_w}{\eta_w}(w_t - \gamma c_t - \varphi n_t) \]
The expectations-adjusted wage-inflation identity is
\(\hat{\pi}^w_t - \hat{\pi}^p_t = w_t - w_{t-1} - \beta \mathbb{E}_t(w_{t+1} - w_t)\).
Combining this identity with the two Phillips curves gives:
\[ w_t - w_{t-1} - \beta \mathbb{E}_t(w_{t+1} - w_t) = -\frac{\psi_w}{\eta_w}(w_t - \gamma y_t - \varphi(y_t - a_t)) + \frac{\psi_p}{\eta_p}(a_t - w_t) \]
The output-gap definition
\(y_t = x_t + \frac{1+\varphi}{\gamma+\varphi}a_t\) implies
\[ \gamma y_t + \varphi(y_t - a_t) = (\gamma+\varphi)x_t + a_t \] so the
real-wage equation can be written as
\[ (1+\beta)w_t - w_{t-1} - \beta\mathbb{E}_t w_{t+1} = -\frac{\psi_w}{\eta_w}(w_t - (\gamma+\varphi)x_t - a_t) + \frac{\psi_p}{\eta_p}(a_t - w_t) \]
Collecting the \(w_t\) terms and using the composite flexibility
parameters \(\kappa_x = \kappa_w(\gamma+\varphi)\) and
\(\kappa_a = \kappa_p + \kappa_w\), along with
\(\Theta_w \equiv 1+\beta+\kappa_a\), yields \[
w_t = \frac{1}{\Theta_w}\left(w_{t-1} + \beta \mathbb{E}_t w_{t+1} + \kappa_x x_t + \kappa_a a_t\right).
\]

\textbf{4. The Macroeconomic Demand Block.} This block derives the TANK
IS curve in output-gap form. The heterogeneity term arises because the
Euler equation applies to saver households, whose consumption differs
from aggregate consumption when \(\sigma_t^c \neq 0\). Combining
\(c^S_t = c_t + \lambda \sigma^c_t\) with the Type S Euler equation
\(r_t = \gamma \mathbb{E}_t \Delta c^S_{t+1}\) gives
\[ r_t = \gamma \mathbb{E}_t(\Delta c_{t+1} + \lambda \Delta \sigma^c_{t+1}) \]
Market clearing, \(c_t = y_t\), and the output-gap relation
\(\Delta y_{t+1} = \Delta x_{t+1} + \frac{1+\varphi}{\gamma+\varphi}\Delta a_{t+1}\)
then give
\[ r_t = \gamma \frac{1+\varphi}{\gamma+\varphi}\mathbb{E}_t\Delta a_{t+1} + \gamma \mathbb{E}_t\Delta x_{t+1} + \gamma\lambda\mathbb{E}_t\Delta \sigma^c_{t+1} \]
Defining the technology/RANK component
\(r_t^R=\gamma\mathbb{E}_t\Delta y_{t+1}^f\) gives (\ref{eq-tank-is}).
Evaluating the same Euler equation at the jointly flexible allocation
gives \(r_t^n=r_t^R+\gamma\lambda\mathbb{E}_t\Delta\sigma_{t+1}^{c,f}\).
Subtracting this flexible-allocation identity yields \[
r_t-r_t^n=\gamma\mathbb{E}_t\Delta x_{t+1}
+\gamma\lambda\mathbb{E}_t\Delta(\sigma_{t+1}^c-\sigma_{t+1}^{c,f}),
\] which rearranges to (\ref{eq-tank-is-natural}). The natural-rate term
and the flexible consumption-gap correction enter together; including
only one would double-count or omit flexible-allocation redistribution.

\subsubsection*{A.2 Own-Lag Adjustment and a Reduced-Form Timing
Diagnostic}\label{a.2-own-lag-adjustment-and-a-reduced-form-timing-diagnostic}
\addcontentsline{toc}{subsubsection}{A.2 Own-Lag Adjustment and a
Reduced-Form Timing Diagnostic}

Own-lag wage adjustment combines an inherited relative-wage state with
responses to expected future wedges. Appendix A.1 gives its reduced
equation: \[
\sigma_t^w=\frac{\sigma_{t-1}^w+\beta \mathbb{E}_t\sigma_{t+1}^w+
\frac{\gamma\kappa_w}{1-\lambda}\omega_t}
{1+\beta+\kappa_w[1+\gamma(\psi_w-1)+\psi_w\varphi]}.
\] For the supplementary diagnostic, define each type's wage change
relative to the aggregate lagged wage as \[
\pi_t^{j,\mathrm{ref}}=w_t^j-w_{t-1}+\pi_t^p.
\] Its population-weighted average equals aggregate wage inflation
\(\pi_t^w\). The diagnostic retains the original forward-looking wage
Phillips curves as reduced-form relations, replacing type-specific
inflation with \(\pi_t^{j,\mathrm{ref}}\) in those relations. Then
\(\pi_t^{H,\mathrm{ref}}-\pi_t^w=-(1-\lambda)\sigma_t^w\), and the
imposed wage-gap equation becomes \[
\sigma_t^w=\frac{\beta \mathbb{E}_t\sigma_{t+1}^w+
\frac{\gamma\kappa_w}{1-\lambda}\omega_t}
{1+\kappa_w[1+\gamma(\psi_w-1)+\psi_w\varphi]}.
\] This equation remains forward looking but has no predetermined
relative-wage state. The change in lag indexing also changes the
contemporaneous coefficient. Comparing its equilibrium solution with the
own-lag model therefore changes the reduced-form forcing map; it does
not isolate a single state while holding that map fixed.

The diagnostic is not the aggregate-lag adjustment-cost model studied in
Miyazaki (2026). In that model, households take the external aggregate
wage reference as given, so their current wage does not enter their
future adjustment-cost denominator. The forward-looking marginal-cost
term is absent, and the relative-wage mapping is static. Retaining the
forward term above is instead an explicit reduced-form choice. Both
constructions preserve type-specific wages and earnings reallocation,
distinguishing them from the common-wage closure below.

\subsubsection*{A.3 The Reduced-Form Common-Wage Contrast and the
Missing Earnings
Buffer}\label{a.3-the-reduced-form-common-wage-contrast-and-the-missing-earnings-buffer}
\addcontentsline{toc}{subsubsection}{A.3 The Reduced-Form Common-Wage
Contrast and the Missing Earnings Buffer}

This subsection gives the algebraic details for the reduced-form
common-wage closure introduced in Section 2. Suppose that wages are
common across household types, so that \(w_t^H = w_t^S = w_t\). The CES
demand schedule then implies \(n_t^H = n_t^S = n_t\). This closure
removes the Type H wage block and the relative labor-demand equation. As
a result, the closure eliminates the labor-reallocation channel
represented by the term \((1-\psi_w)\sigma_t^w\) in Appendix A.1.

The Type H budget constraint and the aggregate resource identity become
\[
c_t^H = w_t + n_t + z_t, \qquad c_t = y_t = a_t + n_t.
\] Hence, \[
c_t - c_t^H = a_t - w_t - z_t = \omega_t.
\] Using \(c_t = \lambda c_t^H + (1-\lambda)c_t^S\), it follows that \[
c_t - c_t^H = (1-\lambda)(c_t^S - c_t^H) = (1-\lambda)\sigma_t^c,
\] so the static consumption-gap relation becomes \[
\sigma_t^c = \frac{\omega_t}{1-\lambda}.
\]

The reduced dynamic system is then \[\begin{aligned}
x_t &= \mathbb{E}_t x_{t+1} - \frac{1}{\gamma}(r_t - r_t^R) - \lambda(\sigma_t^c - \mathbb{E}_t \sigma_{t+1}^c),\\
\pi_t^p& = \beta \mathbb{E}_t \pi_{t+1}^p + \kappa_p(w_t - a_t),\\
r_t &= \phi_\pi \pi_t^p - \mathbb{E}_t \pi_{t+1}^p - m_t,\\
w_t &= \frac{1}{\Theta_w}\left(w_{t-1} + \beta \mathbb{E}_t w_{t+1} + \kappa_x x_t + \kappa_a a_t\right),
\end{aligned}
\] with \(\sigma_t^c = \frac{\omega_t}{1-\lambda}\) and
\(\omega_t = a_t - w_t - z_t\).

By construction, the wage-gap equation disappears. The aggregate real
wage remains sticky and depends on lagged and expected future aggregate
wages. Unless \(\omega_t = 0\), however, the cross-type consumption gap
remains nonzero and the TANK IS curve still differs from its
representative-agent counterpart. This closure is distinct from the
joint flexible-wage and perfect-substitutability limit in the canonical
representation. Here, the labor-reallocation channel is shut down by
assumption; in that joint limit, it remains active and yields
\(\sigma_t^c \to \frac{\varphi}{\varphi+\gamma}\frac{\omega_t}{1-\lambda}\).

\subsection*{Appendix B. Stable Solutions and Equilibrium Impact
Proofs}\label{appendix-b.-stable-solutions-and-equilibrium-impact-proofs}
\addcontentsline{toc}{subsection}{Appendix B. Stable Solutions and
Equilibrium Impact Proofs}

\subsubsection*{B.1 Stable Wage-Gap and Aggregate-Wage
Solutions}\label{b.1-stable-wage-gap-and-aggregate-wage-solutions}
\addcontentsline{toc}{subsubsection}{B.1 Stable Wage-Gap and
Aggregate-Wage Solutions}

\textbf{Proof of Proposition~\ref{prp-stable-wage-gap}.}

The wage-gap equation can be written as a second-order expectational
difference equation:
\[ \beta \mathbb{E}_t\sigma_{t+1}^w - \Theta_\sigma\sigma_t^w + \sigma_{t-1}^w = -\frac{\gamma\kappa_w}{1-\lambda}\omega_t \]
The corresponding characteristic polynomial is
\(\mathfrak p_\sigma(\xi)=\beta\xi^2-\Theta_\sigma\xi+1\).

Saddle-path stability requires one root inside the unit circle and one
outside. Evaluating the polynomial gives \(\mathfrak p_\sigma(0)=1>0\)
and \(\mathfrak p_\sigma(1)=1+\beta-\Theta_\sigma<0\). Since the
polynomial is a convex parabola with \(\beta>0\), it crosses the
horizontal axis exactly once in \((0,1)\). It crosses once more in the
interval \((1, \infty)\).

Let \(\xi_1 \in (0,1)\) denote the stable root and \(\xi_2 > 1\) the
unstable root. Applying standard forward iteration to rule out explosive
paths gives the unique non-explosive solution:
\[ \sigma_t^w = \xi_1 \sigma_{t-1}^w + \frac{\gamma\kappa_w}{(1-\lambda)\beta\xi_2} \sum_{k=0}^{\infty} \left(\frac{1}{\xi_2}\right)^k \mathbb{E}_t \omega_{t+k} \]
The current wage gap therefore depends on both its inherited state
(\(\sigma_{t-1}^w\)) and the expected path of the net distributional
wedge from date \(t\) onward. This proves
Proposition~\ref{prp-stable-wage-gap}.

\textbf{Aggregate wage and wedge.}

Write the aggregate wage equation as \[
\beta\mathbb{E}_tw_{t+1}-\Theta_ww_t+w_{t-1}
=-\mathfrak f_t^w,\qquad
\mathfrak f_t^w=\kappa_xx_t+\kappa_aa_t.
\] The polynomial \(\mathfrak p_w(\chi)=\beta\chi^2-\Theta_w\chi+1\) has
\(\mathfrak p_w(0)=1\) and \(\mathfrak p_w(1)=-\kappa_a<0\). Its roots
satisfy \(0<\chi_1<1<\chi_2\) and \(\beta\chi_1\chi_2=1\). Setting
\(\mathfrak g_t^w=w_t-\chi_1w_{t-1}\) yields \[
\mathbb{E}_t\mathfrak g_{t+1}^w-\chi_2\mathfrak g_t^w
=-\mathfrak f_t^w/\beta.
\] Iteration to horizon \(N\) gives \[
\mathfrak g_t^w
=\frac{1}{\beta\chi_2}
\sum_{k=0}^{N}\chi_2^{-k}\mathbb{E}_t\mathfrak f_{t+k}^w
+\chi_2^{-(N+1)}\mathbb{E}_t\mathfrak g_{t+N+1}^w.
\] Impose \(\lim_{N\to\infty}\chi_2^{-(N+1)}
\mathbb{E}_t\mathfrak g_{t+N+1}^w=0\). For a bounded conditional forcing
path, taking the infinite-horizon limit and using
\(\omega_t=a_t-w_t-z_t\) proves (\ref{eq-equilibrium-wage-wedge}). The
output-gap path remains an equilibrium object throughout this
derivation.

\subsubsection*{B.2 The General Impact Criterion and the AR(1)
Corollary}\label{b.2-the-general-impact-criterion-and-the-ar1-corollary}
\addcontentsline{toc}{subsubsection}{B.2 The General Impact Criterion
and the AR(1) Corollary}

\textbf{Proof of Proposition~\ref{prp-impact-attenuation}.}

When \(\sigma_{t-1}^w=0\), the stable solution implies \[
(1-\psi_w)\sigma_t^w
=-\frac{K_\omega}{(1-\lambda)\beta\xi_2}
\operatorname{PV}_t[\omega].
\] For \(\omega_t\neq0\), dividing the earnings term by direct incidence
and using (\ref{eq-static-dispersion}) yields
(\ref{eq-equilibrium-impact-decomposition}). Opposite signs for the
earnings and direct terms require \(\mathcal B_t^{\mathrm{GE}}>0\). The
net term has the same nonzero sign as direct incidence and a strictly
smaller absolute value exactly when
\(0<1-\mathcal B_t^{\mathrm{GE}}<1\). These requirements are equivalent
to (\ref{eq-equilibrium-offset-condition}), proving
Proposition~\ref{prp-impact-attenuation}. When the inherited gap is
nonzero, retaining the first term of the stable solution gives
(\ref{eq-equilibrium-inherited-decomposition}).

\textbf{An equivalent aggregate-forcing expression.}

Define \(\mathfrak u_t=a_t-z_t\). Substituting
\(w_t=\mathfrak u_t-\omega_t\) into the aggregate wage equation yields
\begin{equation}\protect\phantomsection\label{eq-equilibrium-forced-wedge}{
\begin{aligned}
\beta\mathbb{E}_t\omega_{t+1}-\Theta_w\omega_t+\omega_{t-1}
&=\mathfrak f_t^\omega,\\
\mathfrak f_t^\omega
&\equiv\mathfrak f_t^w+\beta\mathbb{E}_t\mathfrak u_{t+1}
-\Theta_w\mathfrak u_t+\mathfrak u_{t-1}.
\end{aligned}
}\end{equation} Multiply the equation at \(t+k\) by \(\xi_2^{-k}\), take
\(\mathbb{E}_t\), and sum. The tower property gives \[
\begin{aligned}
\beta\xi_2\bigl(\operatorname{PV}_t[\omega]-\omega_t\bigr)
-\Theta_w\operatorname{PV}_t[\omega]
+\omega_{t-1}+\xi_2^{-1}\operatorname{PV}_t[\omega]
=\operatorname{PV}_t[\mathfrak f^\omega].
\end{aligned}
\] Since \(\beta\xi_2+\xi_2^{-1}=\Theta_\sigma\), this becomes
\begin{equation}\protect\phantomsection\label{eq-equilibrium-forcing-pv}{
(\Theta_\sigma-\Theta_w)\operatorname{PV}_t[\omega]
=\operatorname{PV}_t[\mathfrak f^\omega]
+\beta\xi_2\omega_t-\omega_{t-1}.
}\end{equation} If \(\Theta_\sigma\neq\Theta_w\), division by
\(\omega_t\) gives
\begin{equation}\protect\phantomsection\label{eq-equilibrium-forcing-multiplier}{
\mathcal M_t^\omega
=
\frac{\displaystyle\beta\xi_2-\frac{\omega_{t-1}}{\omega_t}
+\frac{\operatorname{PV}_t[\mathfrak f^\omega]}{\omega_t}}
{\Theta_\sigma-\Theta_w}.
}\end{equation} Although this expression is analytic, its forcing still
depends on the full equilibrium paths of output and shocks. The general
condition and resolvent remain valid when \(\Theta_\sigma=\Theta_w\);
only the divided expression is unavailable.

\textbf{Proof of Corollary~\ref{cor-ar1-wedge}.}

Under \(\sigma_{t-1}^w=0\) and
\(\mathbb{E}_t\omega_{t+k}=\rho_\omega^k\omega_t\), the stable solution
above gives \[
\sigma_t^w
=
\frac{\gamma\kappa_w}{(1-\lambda)\beta\xi_2}
\sum_{k=0}^{\infty}
\left(\frac{\rho_\omega}{\xi_2}\right)^k\omega_t
=
\frac{\gamma\kappa_w}{(1-\lambda)\beta(\xi_2-\rho_\omega)}\omega_t.
\] Substituting this expression into (\ref{eq-static-dispersion}) gives
(\ref{eq-geometric-offset-share}). The offset share is positive because
\(\psi_w>1\), \(\gamma>0\), \(\kappa_w>0\), and \(\xi_2>1>\rho_\omega\).
To establish the strict upper bound, use
\(\beta\xi_2=\Theta_\sigma-\xi_2^{-1}\) and the definition of
\(\Theta_\sigma\) to obtain \[
\begin{aligned}
&\beta(\xi_2-\rho_\omega)-(\psi_w-1)\gamma\kappa_w\\
&\qquad=
(1-\xi_2^{-1})
+\beta(1-\rho_\omega)
+\kappa_w(1+\psi_w\varphi)
>0.
\end{aligned}
\] Hence \(0<\mathcal{B}(\rho_\omega)<1\). Applying
Proposition~\ref{prp-impact-attenuation} then establishes
Corollary~\ref{cor-ar1-wedge}.

\subsubsection*{B.3 Exact Verification of the Benchmark Equilibrium
Impact}\label{b.3-exact-verification-of-the-benchmark-equilibrium-impact}
\addcontentsline{toc}{subsubsection}{B.3 Exact Verification of the
Benchmark Equilibrium Impact}

This subsection verifies the condition in
Proposition~\ref{prp-impact-attenuation} separately for each benchmark
innovation from the steady state. The benchmark parameter values are
rational numbers. After clearing the timing denominators in the
canonical system's characteristic determinant, its nonzero
characteristic roots satisfy \[
(\zeta-\rho_\star)^3\mathfrak p_{\mathrm{GE}}(\zeta)=0,
\qquad \rho_\star=\frac45,
\] where
\begin{equation}\protect\phantomsection\label{eq-equilibrium-characteristic-polynomial}{
\begin{aligned}
\mathfrak p_{\mathrm{GE}}(\zeta)
={}&2910897\zeta^6-30098871\zeta^5+98658351\zeta^4\\
&-144731677\zeta^3+101811300\zeta^2
-31070000\zeta+3000000.
\end{aligned}
}\end{equation} Let \(\mathcal V_{\mathrm{St}}(\zeta)\) denote the
sign-variation count in its Sturm sequence, omitting zero entries. Exact
rational evaluation gives \(\mathcal V_{\mathrm{St}}(0)=5\) and
\(\mathcal V_{\mathrm{St}}(1)=3\). The following intervals isolate the
two endogenous roots in \((0,1)\):
\begin{equation}\protect\phantomsection\label{eq-equilibrium-root-intervals}{
\begin{aligned}
\frac{1739303033}{10^{10}}&<\zeta_-<
\frac{1739303034}{10^{10}},\\
\frac{4641738104}{10^{10}}&<\zeta_+<
\frac{4641738105}{10^{10}}.
\end{aligned}
}\end{equation} The Sturm variation count falls from \(5\) to \(4\)
across the first interval and from \(4\) to \(3\) across the second. For
the transformed polynomial \[
\widetilde{\mathfrak p}_{\mathrm{GE}}(\upsilon)
=(1-\upsilon)^6
\mathfrak p_{\mathrm{GE}}\left(\frac{1+\upsilon}{1-\upsilon}\right),
\] the first-column signs of the exact Routh table are
\(+,-,+,+,+,-,+\). By the Routh criterion, the four sign changes imply
that the remaining four roots lie outside the unit circle. Thus the
stable solution has the common external root \(\rho_\star\) and two
distinct endogenous roots \(\zeta_+,\zeta_-\).

Normalize an endogenous homogeneous mode to \(w=1\) and \(\omega=-1\).
The wage-gap equation then gives its relative-wage loading \[
\mathsf L_\sigma(\zeta)
=\frac{80\zeta}{3(99\zeta^2-599\zeta+100)}.
\] For a unit innovation of type \(i\),
Table~\ref{tbl-equilibrium-particular} gives the coefficients of the
particular solution at \(\rho_\star\).

\begin{longtable}[]{@{}
  >{\raggedright\arraybackslash}p{(\linewidth - 6\tabcolsep) * \real{0.2500}}
  >{\raggedleft\arraybackslash}p{(\linewidth - 6\tabcolsep) * \real{0.2500}}
  >{\raggedleft\arraybackslash}p{(\linewidth - 6\tabcolsep) * \real{0.2500}}
  >{\raggedleft\arraybackslash}p{(\linewidth - 6\tabcolsep) * \real{0.2500}}@{}}
\caption{External-mode
coefficients}\label{tbl-equilibrium-particular}\tabularnewline
\toprule\noalign{}
\begin{minipage}[b]{\linewidth}\raggedright
Innovation
\end{minipage} & \begin{minipage}[b]{\linewidth}\raggedleft
\(\vartheta_{\star i}^w\)
\end{minipage} & \begin{minipage}[b]{\linewidth}\raggedleft
\(\vartheta_{\star i}^{\sigma^w}\)
\end{minipage} & \begin{minipage}[b]{\linewidth}\raggedleft
\(\vartheta_{\star i}^\omega\)
\end{minipage} \\
\midrule\noalign{}
\endfirsthead
\toprule\noalign{}
\begin{minipage}[b]{\linewidth}\raggedright
Innovation
\end{minipage} & \begin{minipage}[b]{\linewidth}\raggedleft
\(\vartheta_{\star i}^w\)
\end{minipage} & \begin{minipage}[b]{\linewidth}\raggedleft
\(\vartheta_{\star i}^{\sigma^w}\)
\end{minipage} & \begin{minipage}[b]{\linewidth}\raggedleft
\(\vartheta_{\star i}^\omega\)
\end{minipage} \\
\midrule\noalign{}
\endhead
\bottomrule\noalign{}
\endlastfoot
Technology & \(4085425/5201722\) & \(37700/2600861\) &
\(1116297/5201722\) \\
Monetary & \(3207750/2600861\) & \(-650000/7802583\) &
\(-3207750/2600861\) \\
Transfer & \(116350/2600861\) & \(-550600/7802583\) &
\(-2717211/2600861\) \\
\end{longtable}

\emph{Notes:} Each column is the coefficient on \(\rho_\star^h\) in the
corresponding variable's particular solution, normalized to a unit
innovation. These coefficients solve the benchmark canonical equations
at \(\rho_\star\); they are not the total impact responses.

Set \(\mathsf L_\pm^\sigma=\mathsf L_\sigma(\zeta_\pm)\). Setting the
inherited aggregate wage and wage gap to zero gives the endogenous wedge
loadings
\begin{equation}\protect\phantomsection\label{eq-equilibrium-modal-loadings}{
\begin{aligned}
\vartheta_{+i}^\omega
&=\frac{\zeta_+
(\vartheta_{\star i}^{\sigma^w}
-\mathsf L_-^\sigma\vartheta_{\star i}^w)}
{\rho_\star(\mathsf L_+^\sigma-\mathsf L_-^\sigma)},\\
\vartheta_{-i}^\omega
&=\frac{\zeta_-
(\mathsf L_+^\sigma\vartheta_{\star i}^w
-\vartheta_{\star i}^{\sigma^w})}
{\rho_\star(\mathsf L_+^\sigma-\mathsf L_-^\sigma)}.
\end{aligned}
}\end{equation} Together with the external loading, these coefficients
establish (\ref{eq-equilibrium-wedge-modes}). All three loadings are
nonzero for each innovation. Summing the resulting geometric paths
yields exactly (\ref{eq-equilibrium-modal-multiplier}).

At this calibration, \[
\xi_2=\frac{599+\sqrt{319201}}{198},
\qquad
\frac{58786806094}{10^{10}}<\xi_2<
\frac{58786806095}{10^{10}},
\qquad
\frac{K_\omega}{\beta\xi_2}
=\frac{360}{599+\sqrt{319201}}.
\] Substituting these root intervals into
(\ref{eq-equilibrium-modal-loadings}) and
(\ref{eq-equilibrium-modal-multiplier}) using rational interval
arithmetic gives
\begin{equation}\protect\phantomsection\label{eq-equilibrium-certified-offsets}{
\begin{aligned}
\frac{342940}{10^6}
&<\mathcal B_{0,a}^{\mathrm{GE}}<\frac{342941}{10^6},\\
\frac{358490}{10^6}
&<\mathcal B_{0,z}^{\mathrm{GE}}<\frac{358491}{10^6},\\
\frac{388804}{10^6}
&<\mathcal B_{0,m}^{\mathrm{GE}}<\frac{388805}{10^6}.
\end{aligned}
}\end{equation} The denominators and the signs of the nonzero loadings
are also certified over these intervals. Every bound lies strictly
within \((0,1)\), proving the benchmark impact result without using a
finite-horizon approximation or relying on rounded impulse responses. As
a separate numerical check, the modal solution reproduces the saved
first-order equilibrium solution.

The result applies to individual innovations, not arbitrary simultaneous
mixtures. To see why, retain zero inherited states and define the impact
direction \[
\mathbf n=
\frac{\mathbf b_a}{\mathbf v_\omega'\mathbf b_a}
-\frac{\mathbf b_m}{\mathbf v_\omega'\mathbf b_m}.
\] Then \(\mathbf v_\omega'\mathbf n=0\), whereas \[
\mathbf v_\omega'(I-\mathbf T_{\mathrm{eq}}/\xi_2)^{-1}\mathbf n
=\mathcal M_{0,a}^\omega-\mathcal M_{0,m}^\omega\neq0.
\] Adding any multiple of \(\mathbf n\) to a nonzero-wedge impact
direction leaves the current wedge unchanged but changes its discounted
expected sum. Consequently, there is no uniform result governing both
the sign and the upper bound across all simultaneous signed shock
mixtures. Each mixture can be scaled arbitrarily close to the steady
state without changing this ratio, so the qualification is consistent
with local linear analysis.

\subsection*{Appendix C. Numerical Diagnostics and Additional
Quantitative
Comparisons}\label{appendix-c.-numerical-diagnostics-and-additional-quantitative-comparisons}
\addcontentsline{toc}{subsection}{Appendix C. Numerical Diagnostics and
Additional Quantitative Comparisons}

\subsubsection*{C.1 Numerical Local Determinacy of the Full Canonical
System}\label{c.1-numerical-local-determinacy-of-the-full-canonical-system}
\addcontentsline{toc}{subsubsection}{C.1 Numerical Local Determinacy of
the Full Canonical System}

At the baseline calibration, the full system under passive transfers is
determinate. The numerical root count yields four unstable roots, five
stable roots, and no unit roots. The largest stable root is \(0.8\),
while the smallest unstable root is approximately \(1.13\).

The passive-transfer calculations use the first-order companion form of
the same linearized canonical system implemented in the quantitative
code and apply the Blanchard--Kahn root count. A point is determinate
when the number of roots outside the unit circle equals the number of
forward-looking variables. A smaller count implies indeterminacy; a
larger count implies no stable solution.

\begin{figure}[H]

\includegraphics[width=1\linewidth,height=\textheight,keepaspectratio]{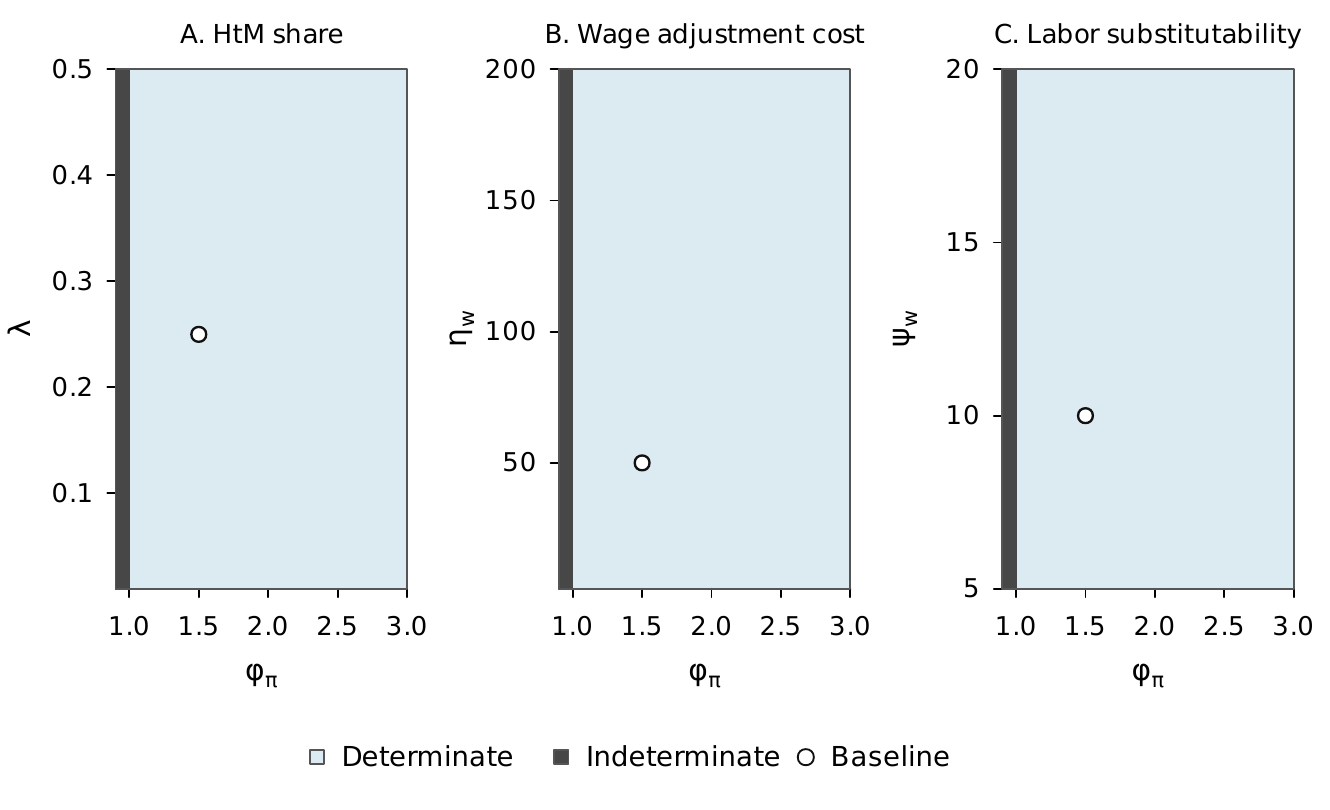}

\par\smallskip
\begin{minipage}{0.98\linewidth}
\raggedright
\footnotesize \textit{Notes:} Panels vary $(\phi_\pi,\lambda)$, $(\phi_\pi,\eta_w)$, and $(\phi_\pi,\psi_w)$ in the passive-transfer canonical system. The white marker denotes the baseline calibration $(\phi_\pi,\lambda,\eta_w,\psi_w)=(1.5,0.25,50,10)$. Classification uses the Blanchard--Kahn root count at sampled grid nodes; shaded cells display the nearest node and do not locate an analytical boundary.
\end{minipage}

\caption{\label{fig-appendix-determinacy}Local-determinacy maps}

\end{figure}

Figure~\ref{fig-appendix-determinacy} reports three parameter planes,
each with the Taylor-rule inflation coefficient \(\phi_\pi\) on the
horizontal axis. The figure shows the \((\phi_\pi,\lambda)\),
\((\phi_\pi,\eta_w)\), and \((\phi_\pi,\psi_w)\) planes; the baseline
point lies within a determinate region in all three. Across the
implemented grids, all points with \(\phi_\pi=0.90\) or
\(\phi_\pi=0.97\) are indeterminate over the full ranges of
\(\lambda \in [0.01,0.50]\), \(\eta_w \in [2,200]\), and
\(\psi_w \in [5,20]\), whereas all points with \(\phi_\pi \ge 1.04\) are
determinate. Within these baseline-centered, one-parameter-at-a-time
slices, this pattern is consistent with the Taylor-principle threshold:
the classification changes only near \(\phi_\pi\approx1\).\footnote{
  These slices do not imply that wage rigidity is irrelevant for
  determinacy. In a supplementary check at
  \((\lambda,\psi_w,\phi_\pi)=(0.70,10,1.5)\), the full own-lag system
  is indeterminate at \(\eta_w=2\) (three unstable roots for four
  forward-looking variables) but determinate at \(\eta_w=10\) and \(50\)
  (four unstable roots). The projection of the endogenous stable
  eigenspace onto the two lagged wage states has full rank at all three
  points. Thus, wage rigidity can affect determinacy when the
  hand-to-mouth share and the wage-adjustment cost are varied jointly.}
Additional grids for \((\phi_\pi,\eta_p)\) and \((\phi_\pi,\psi_p)\)
display the same pattern and are therefore omitted from the figure.

\subsubsection*{C.2 Cross-Shock Distributional
Accounting}\label{c.2-cross-shock-distributional-accounting}
\addcontentsline{toc}{subsubsection}{C.2 Cross-Shock Distributional
Accounting}

The decomposition in (\ref{eq-static-dispersion}) is an accounting
identity, so its form does not depend on the shock.
Table~\ref{tbl-cross-shock-accounting} applies the same decomposition to
one-time 1 percent monetary, transfer, and technology innovations. At
the date of the maximum absolute consumption gap, earnings reallocation
offsets 42.2, 35.8, and 34.3 percent of direct incidence, respectively.
The corresponding 20-quarter offsets are 43.9, 43.2, and 42.0 percent.

\begin{table}[H]

\caption{\label{tbl-cross-shock-accounting}Cross-shock accounting}

\centering{

\centering
\begin{threeparttable}
\begin{tabular}[t]{lrr}
\toprule
Shock & Peak offset (\%) & 20-quarter offset (\%)\\
\midrule
Monetary policy & 42.2 & 43.9\\
Transfer & 35.8 & 43.2\\
Technology & 34.3 & 42.0\\
\bottomrule
\end{tabular}
\begin{tablenotes}
\item \textit{Notes:} The peak offset is the negative of the ratio of the earnings term to direct incidence at the date maximizing $|\sigma_t^c|$. Over horizons $h=0,\ldots,19$, the cumulative offset is $1-\sum_{h=0}^{19}|\sigma_h^c|/\sum_{h=0}^{19}|\omega_h/(1-\lambda)|$. Both compare the direct and net terms along the realized type-specific-wage equilibrium path; they are accounting offsets, not general-equilibrium reductions relative to a common-wage economy.
\end{tablenotes}
\end{threeparttable}

}

\end{table}%

The technology experiment reverses the incidence pattern. At the date of
the maximum absolute consumption gap, \[
\underbrace{+0.009034}_{\text{direct profit-transfer incidence}}
+
\underbrace{(-0.003098)}_{\text{earnings reallocation}}
=
\underbrace{+0.005936}_{\text{net consumption gap}}.
\] Direct incidence is positive and the earnings term is negative,
reversing the pattern in the monetary and transfer experiments. At every
date with a nonzero response, the earnings term and direct incidence
have opposite signs in all three experiments. Type-specific labor
reallocation therefore acts as a buffer under either sign of direct
incidence rather than as a one-way transfer.

\begin{figure}[H]

\centering{

\includegraphics[width=1\linewidth,height=\textheight,keepaspectratio]{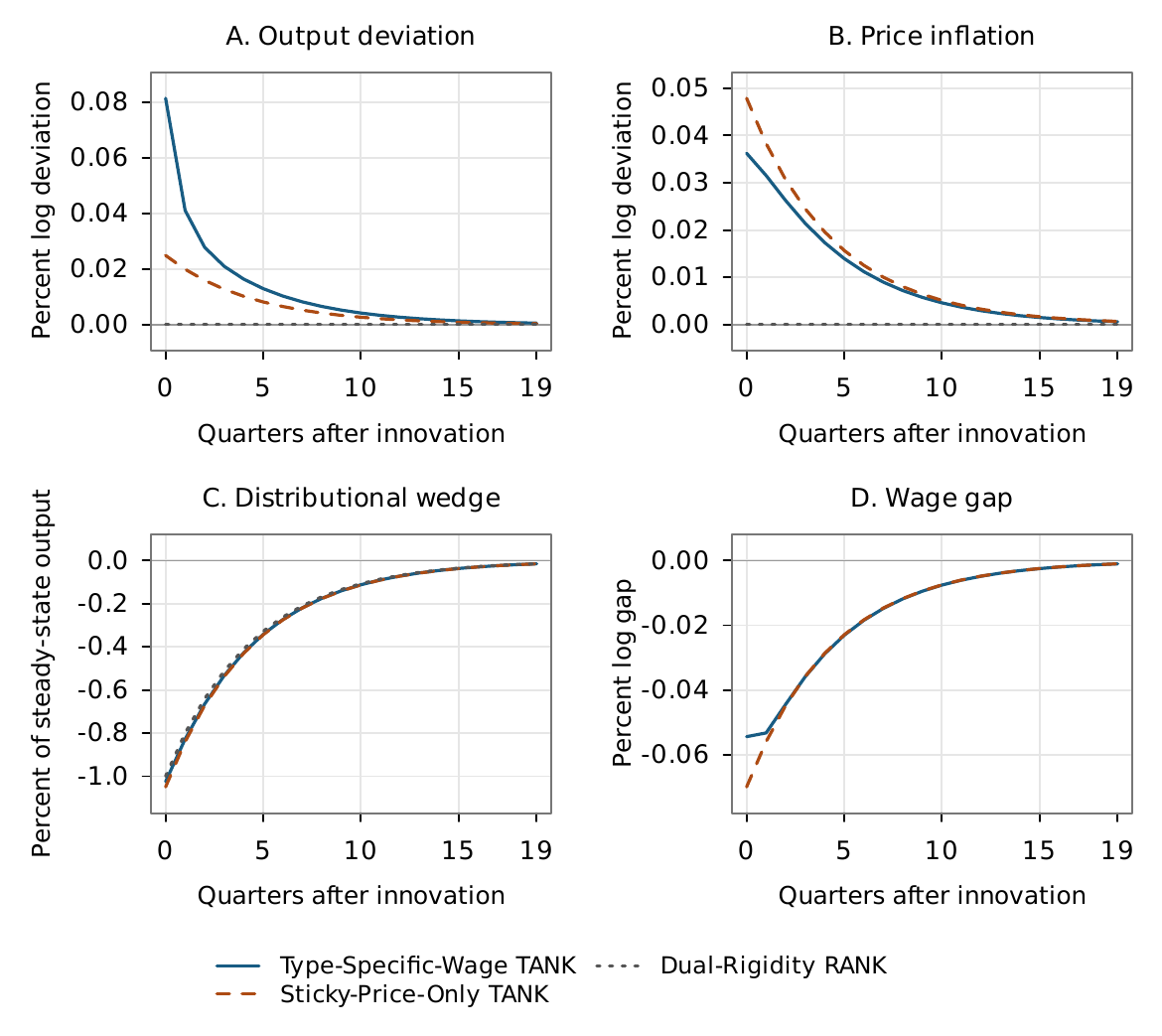}

\par\smallskip
\begin{minipage}{0.98\linewidth}
\raggedright
\footnotesize \textit{Notes:} Deterministic first-order responses to a transfer innovation of $e_{z,0}=0.01$ per Type H household, equal to 0.25 percent of steady-state output in aggregate at $\lambda=0.25$, over 20 quarters. Panels (A)--(C) report $x_t$, $\pi_t^p$, and $\omega_t$ for Type-Specific-Wage TANK, Sticky-Price-Only TANK, and Dual-Rigidity RANK. Panel (D) reports $\sigma_t^w$ for the two heterogeneous economies. Responses are percent log deviations, while the wedge is in percent of steady-state output.
\end{minipage}

}

\caption{\label{fig-transfer-shock}Transfer-shock responses}

\end{figure}

The transfer experiment also illustrates how sticky and flexible
type-specific wages differ in their aggregate and wage-gap dynamics.
Figure~\ref{fig-transfer-shock} reports the underlying responses. The
peak output-gap response is \(8.14\times 10^{-4}\) in Type-Specific-Wage
TANK, compared with \(2.49\times 10^{-4}\) in Sticky-Price-Only TANK,
for a ratio of about \(3.27\) under the benchmark calibration. The
wage-gap response in Type-Specific-Wage TANK has a half-life of five
quarters, compared with four quarters under flexible wages. The
aggregate ordering is therefore not uniform across shocks, metrics, and
horizons; the monetary experiment with the reduced-form common-wage
closure remains the paper's main cross-closure contrast.

\subsubsection*{C.3 Sensitivity of the Quantitative
Results}\label{c.3-sensitivity-of-the-quantitative-results}
\addcontentsline{toc}{subsubsection}{C.3 Sensitivity of the Quantitative
Results}

The joint parameter grid reported in Section 4 combines
\(\lambda\in\{0.10,0.25,0.40\}\), \(\psi_w\in\{2,5,10,20\}\), and
\(\eta_w\in\{25,50,100\}\). The model is solved under own-lag
adjustment, the reduced-form timing diagnostic in Appendix A.2, and the
reduced-form common-wage closure at each of the 36 cells; all three are
determinate throughout the grid. The replication materials report the
joint-grid results and the state/forcing decomposition.

This subsection reports baseline-centered one-parameter-at-a-time
sensitivity checks for the main quantitative magnitudes in Section 4.
Each sweep keeps the remaining parameters at their baseline values and
evaluates a 0.01 innovation over 20 quarters. For transfers this fixes
the payment per Type H household, so aggregate transfers vary with the
Type H share. The varied parameters are
\(\lambda \in \{0.05,0.15,0.25,0.35,0.45\}\),
\(\eta_w \in \{10,25,50,75,100,150\}\), and
\(\psi_w \in \{5,7.5,10,12.5,15,20\}\). All parameter points reported
below remain determinate.

\begin{figure}[H]

\includegraphics[width=1\linewidth,height=\textheight,keepaspectratio]{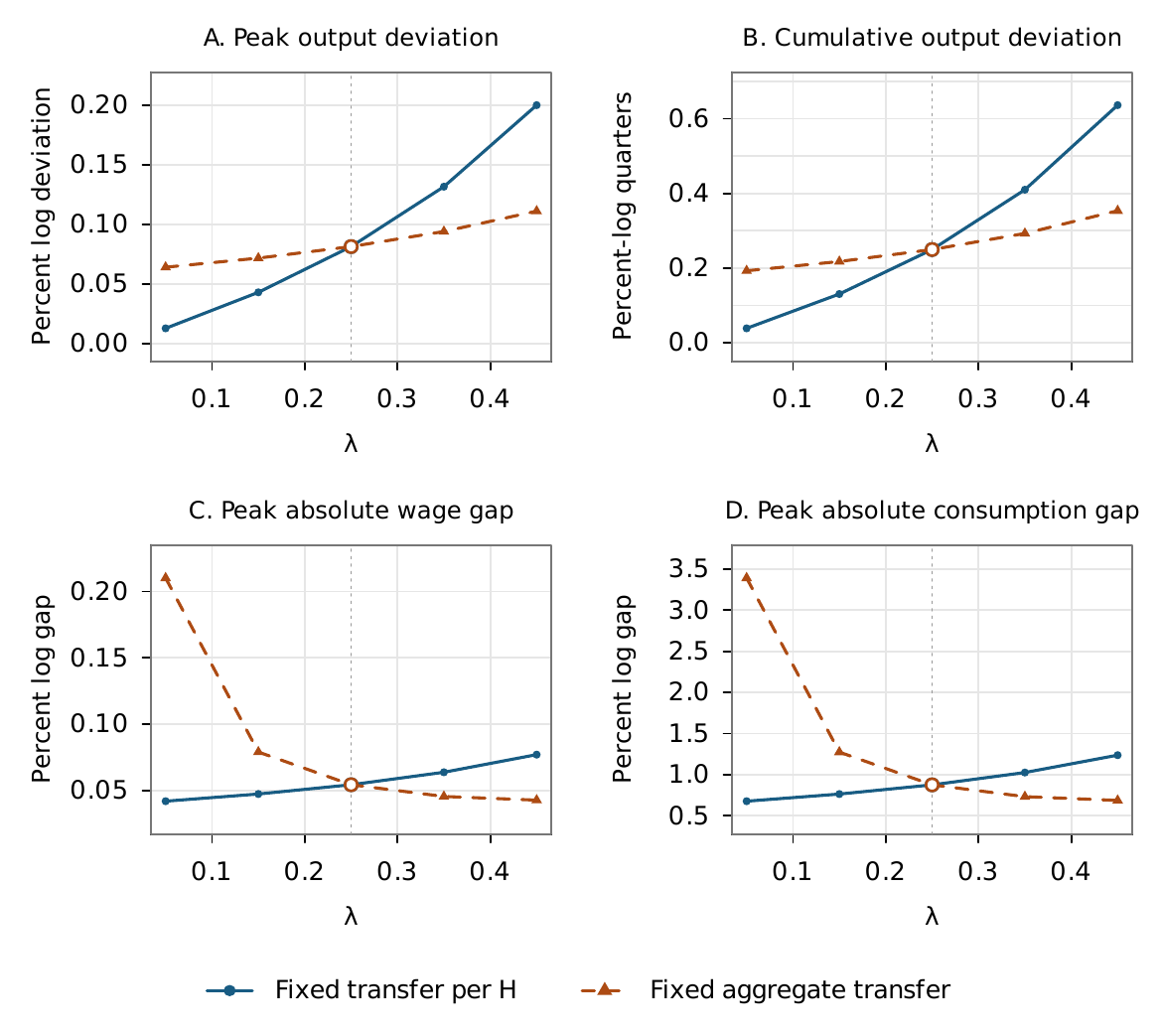}

\par\smallskip
\begin{minipage}{0.98\linewidth}
\raggedright
\footnotesize \textit{Notes:} First-order transfer responses over $h=0,\ldots,19$. The fixed-per-recipient experiment sets $e_{z,0}=0.01$; the fixed-total experiment sets $\lambda e_{z,0}=0.0025$. All other parameters remain at baseline. Responses are percent log deviations and cumulative output is in percent-log quarters. The different consumption-gap slopes combine changing household shares with the stated fiscal normalization.
\end{minipage}

\caption{\label{fig-transfer-normalization}Transfer normalization and
household shares}

\end{figure}

The sensitivity checks deliver two main patterns. Under a fixed
per-recipient transfer, higher hand-to-mouth shares raise the peak
output response and widen the cross-type consumption gap. This
comparison also raises the aggregate transfer amount. Under a fixed
total transfer, the peak output response still rises, but the maximum
consumption gap falls; these normalizations therefore have different
distributional comparative statics. For monetary shocks, higher
wage-adjustment costs raise peak and cumulative output responses but
lower the distributional metrics, changing both the size and persistence
of the wage-gap path. The online appendix reports the full sweeps in
separate figures for each parameter and a selected-value table.
Figure~\ref{fig-transfer-normalization} compares the two fiscal
normalizations directly.

The fixed-total experiment sets \(\lambda e_{z,0}=0.0025\) at every
population share, matching the aggregate transfer at the benchmark. Thus
\(e_{z,0}=0.0025/\lambda\). Because the model is linear, the same
experiment can also be recovered by multiplying the fixed-per-recipient
responses at each parameter value by \(0.25/\lambda\); direct solutions
verify this scaling.

At \(\lambda=0.05\) and \(0.45\), the fixed-per-recipient maximum
consumption gaps are 0.678 and 1.238 log percent, whereas the
fixed-total gaps are 3.392 and 0.688 log percent. The fixed-total peak
output response rises from 0.064 to 0.111 log percent, compared with
0.013 to 0.200 under the original normalization. Statements about
increasing amplification in \(\lambda\) must therefore specify whether
transfer expenditure or the individual payment is held fixed.

\end{document}